\documentclass[prd,twocolumn,reprint,preprintnumbers,nofootinbib,superscriptaddress]{revtex4-1}
\input{header.tex}

\begin{document}

\reportnum{-2}{CERN-TH-2024-189}

\title{Dynamics of metastable Standard Model particles from long-lived particle decays in the MeV primordial plasma}

\author{Kensuke~Akita}
\email{kensuke@hep-th.phys.s.u-tokyo.ac.jp}
\affiliation{Department of Physics, The University of Tokyo, Bunkyo-ku, Tokyo 113-0033, Japan}
\author{Gideon~Baur}
\email{gbaur@uni-bonn.de}
\affiliation{Institut für Astroteilchen Physik, Karlsruher Institut für Technologie (KIT), Hermann-von-Helmholtz-Platz 1, 76344 Eggenstein-Leopoldshafen, Germany}
\affiliation{Bethe Center for Theoretical Physics, Universität Bonn, D-53115, Germany}
\author{Maksym~Ovchynnikov}
\email{maksym.ovchynnikov@cern.ch}
\affiliation{Theoretical Physics Department, CERN, 1211 Geneva 23, Switzerland}
\affiliation{Institut für Astroteilchen Physik, Karlsruher Institut für Technologie (KIT), Hermann-von-Helmholtz-Platz 1, 76344 Eggenstein-Leopoldshafen, Germany}
\author{Thomas Schwetz}
\email{schwetz@kit.edu}
\affiliation{Institut für Astroteilchen Physik, Karlsruher Institut für Technologie (KIT), Hermann-von-Helmholtz-Platz 1, 76344 Eggenstein-Leopoldshafen, Germany}
\author{Vsevolod Syvolap}
\email{sivolapseva@gmail.com}
\affiliation{Instituut-Lorentz, Leiden University, Niels Bohrweg 2, 2333 CA Leiden, The Netherlands}

\date{\today}

\begin{abstract}
We investigate the cosmological impact of hypothetical unstable new physics particles that decay in the MeV-scale plasma of the Early Universe. Focusing on scenarios where the decays produce metastable species such as muons, pions, and kaons, we systematically analyze the dynamics of these particles using coupled Boltzmann equations governing their abundances. Our results demonstrate that the metastable species can efficiently annihilate or interact with nucleons, often leading to their disappearance before decay. The suppression of decay significantly alters the properties of cosmic neutrinos, impacting cosmological observables like Big Bang nucleosynthesis and the Cosmic Microwave Background. To support further studies, we provide two public codes: the \texttt{Mathematica} code that traces the evolution of these metastable particles, as well as the \texttt{python}-based unintegrated neutrino Boltzmann solver that uses this evolution as an input and may be applied to a broad range of scenarios. We then utilize them for studying a few particular new physics models.
\end{abstract}

\maketitle

\tableofcontents

\section{Introduction and summary}
\label{sec:introduction}

The thermal plasma of the Early Universe near the epoch of neutrino decoupling, at temperatures $T\lesssim 5\text{ MeV}$~\cite{Dolgov:2002wy}, serves as a crucial window into potential new physics. Any new particles or interactions present during this period can leave imprints on primordial neutrinos, affecting their abundance and energy distribution. These modifications, in turn, influence key cosmological observables, including primordial nuclear abundances~\cite{Sarkar:1995dd,Dolgov:2000jw,Kohri:2001jx,Hannestad:2004px,Pospelov:2010hj,Kawasaki:2017bqm,Boyarsky:2020dzc}, the Cosmic Microwave Background (CMB)~\cite{Ellis:1984eq,Moroi:1993mb,Kawasaki:1994af,Giudice:2000ex,Hannestad:2004px,Kanzaki:2007pd,Fradette:2017sdd,Fradette:2018hhl,Hasegawa:2019jsa,Sabti:2020yrt,Boyarsky:2021yoh,Mastrototaro:2021wzl,Rasmussen:2021kbf}, and the cosmological implications of neutrino mass~\cite{Alvey:2021sji,Alvey:2021xmq,Escudero:2022gez,Naredo-Tuero:2024sgf}. 

One intriguing scenario involves the existence of hypothetical Long-Lived Particles (LLPs), $X$, with lifetimes $\tau_{X} \lesssim 1\,\text{s}$. These particles can decay into Standard Model (SM) species, such as neutrinos, nucleons, electromagnetic (EM) particles -- $e^{\pm}$ and photons, and various \textit{metastable particles}
\begin{equation}
    Y,\bar Y = \mu^\pm,\pi^\pm, K^\pm, K_L\,. 
\end{equation}
The specific decay channels determine how LLPs influence the neutrino population and primordial nuclear abundances. A key quantity in respect of neutrinos is the effective number of neutrino species $N_{\rm eff}$, showing how much energy is stored in the neutrino sector per the energy of the EM plasma. In the absence of other relativistic particles beyond the SM, $N_{\rm eff}$ is defined as the properly weighted ratio of neutrino and photon energy densities (see later for a precise definition).

Decays into EM particles can heat the EM plasma, thereby reducing $N_{\text{eff}}$. On the other hand, decays into neutrinos have effects that depend on the energy of the injected neutrinos, $E_{\nu}$, relative to the thermal neutrino energy $\sim 3T$. 
If $E_{\nu}\simeq 3T$, decays heat the neutrino plasma, which leads to an increase of $N_{\text{eff}}$ without substantial neutrino spectral distortions~\cite{EscuderoAbenza:2020cmq}.
However, if $E_{\nu} \gg 3T$, $N_{\text{eff}}$ can also decrease because of non-trivial effects related to non-thermal distortions of the neutrino momentum distributions~\cite{Boyarsky:2021yoh,Rasmussen:2021kbf,Ovchynnikov:2024rfu,Ovchynnikov:2024xyd}. 
Examples of LLPs that decay into $Y$ particles include Higgs-like scalars~\cite{Boiarska:2019jym}, generic pseudoscalars such as axion-like particles with various coupling schemes~\cite{Beacham:2019nyx,Bauer:2020jbp,Bauer:2021mvw,Aloni:2018vki,DallaValleGarcia:2023xhh}, particles coupled to quark currents like dark photons and $B-L$ mediators~\cite{Ilten:2018crw}, Heavy Neutral Leptons~\cite{Bondarenko:2018ptm}, and neutralinos. 

The metastable particles $Y$ may subsequently decay into neutrinos, other $Y$\!s, and EM particles. As the decays are governed by weak interactions, the $Y$\!s' inverse lifetimes are relatively low, $\tau_{Y}^{-1} \sim (10^{6}-10^{8})\,\text{s}^{-1}$, exceeding the characteristic interaction rates of $Y$ with the primordial plasma. Consequently, $Y$ particles can engage in various processes before decaying. For charged $Y$ particles at temperatures $T \gtrsim 1\text{ keV}$, frequent interactions with electrons and photons transfer their kinetic energy to the EM plasma~\cite{Reno:1987qw,Kohri:2001jx,Pospelov:2010cw,Fradette:2017sdd}. This was incorporated in the studies~\cite{Fradette:2017sdd,Fradette:2018hhl,Gelmini:2020ekg,Sabti:2020yrt,Boyarsky:2021yoh,Rasmussen:2021kbf}, which examined the impact of LLP decays into $Y$ particles on neutrino properties. These works generally assumed that $Y$ particles inevitably decay.

In this paper, we highlight critical aspects of $Y$ particles dynamics that have been overlooked. Specifically, before decaying, they can undergo processes that lead to their disappearance without producing neutrinos. These processes include annihilations $Y+\bar{Y}\to \text{SM}$, where the antiparticle $\bar{Y}$ is similarly produced by the decaying LLP, and interactions with nucleons $Y+\mathcal{N}\to \mathcal{N}' + \text{SM}$.\footnote{The meson-driven $p\leftrightarrow n$ processes have been considered in the works~\cite{Reno:1987qw,Kohri:2001jx,Kawasaki:2004qu,Pospelov:2010cw,Kawasaki:2017bqm,Hasegawa:2019jsa,Boyarsky:2020dzc} in the context of the impact of new physics on primordial nuclear abundances. However, to the best of our knowledge, they have not been included in any previous study of the impact on neutrinos.} Although the instant abundances of $Y$, $\bar{Y}$, and nucleons are small, the large interaction cross-sections mediated by strong or electromagnetic forces render these processes highly efficient. Depending on the temperature, their rates can compete or even significantly exceed the decay rate, potentially preventing any neutrino injection. Consequently, the properties of cosmic neutrinos are significantly altered compared to scenarios where $Y$ decays are inevitable. We provide a public \texttt{Mathematica} code, which allows us to follow the actual evolution of $Y$ particles as well as nucleon abundances in the Early Universe\footnote{Available on \faGithub \cite{GitHub-metastable} and Zenodo~\cite{Zenodo}.}. As a result, we find that
within the lifetime range $0.01\,\text{s} \lesssim \tau_{X}\lesssim 10\,\text{s}$, the effective number of relativistic degrees of freedom $N_{\text{eff}}$ and the degree of neutrino spectral distortions are substantially reduced, while in presence of the charged kaons the energy distributions of neutrinos and antineutrinos may become asymmetric.

To study the impact of the dynamics of metastable particles on the cosmic neutrinos, we have developed a Boltzmann solver for the momentum-dependent neutrino distribution functions, based on ref.~\cite{Akita:2020szl}.\footnote{Available on \faGithub \cite{GitHub-solver}.} The code allows us to calculate the evolution of neutrinos in the presence of a broad range of new physics particles such as Higgs-like scalars, Majorons, dark photons, ALPs, and electromagnetically decaying relics. The non-trivial dynamics of intermediate metastable SM particles is considered in a self-consistent way. Hence, our work extends previous studies, see, e.g.,~\cite{Dolgov:2000jw,Kawasaki:2004qu,Hannestad:2004px,Ruchayskiy:2012si,Kawasaki:2017bqm,Gariazzo:2019gyi,Sabti:2020yrt,Mastrototaro:2021wzl,Rasmussen:2021kbf}, and offers a flexible public tool to be used by the community. As we will show below, in light of future precise CMB measurements, the unintegrated Boltzmann approach is needed to predict $N_{\text{eff}}$ even for the simplest scenarios, such as LLPs decaying purely electromagnetically. In particular, it should replace various approximate approaches to solve the neutrino Boltzmann equations (see, e.g.,~\cite{Fradette:2017sdd,Fradette:2018hhl,Boyarsky:2020dzc,Chao:2022mcu,Chen:2024cla,Chang:2024mvg}), which may not be accurate enough. 

The paper is organized as follows. In Sec.~\ref{sec:Y}, we describe the properties of the metastable particles $Y$ and their interactions in the primordial plasma. Sec.~\ref{sec:approach} discusses our two-step scheme to calculate the metastable particles and the evolution of neutrinos. Sec.~\ref{sec:metastable-dynamics} describes how we calculate the dynamics of $Y$ particles and nucleon densities. In particular, in Subsection~\ref{sec:simple}, we conduct a simplified analysis for the cases of muons and charged pions, illustrating that they prefer to disappear before decaying at MeV temperatures. In Sec.~\ref{sec:unintegrated}, we describe our numeric approach to solve the neutrino Boltzmann equations in the presence of metastable particles. Sec.~\ref{sec:case-studies-qualitative} contains a qualitative discussion on how the dynamics of $Y$\!s influences neutrino properties, including $N_{\text{eff}}$, the neutrino distribution function, and the neutrino-antineutrino energy asymmetry. Sec.~\ref{sec:case-studies} explores a few models with LLPs, such as Higgs-like scalars and Heavy Neutral Leptons, and analyzes how they affect the neutrino population based on the methods from the previous sections. Finally, we conclude in Sec.~\ref{sec:conclusions}.

\section{Interactions of metastable particles in the primordial MeV plasma}
\label{sec:Y}

\begin{figure*}[t!]
    \centering
    \includegraphics[width=\linewidth]{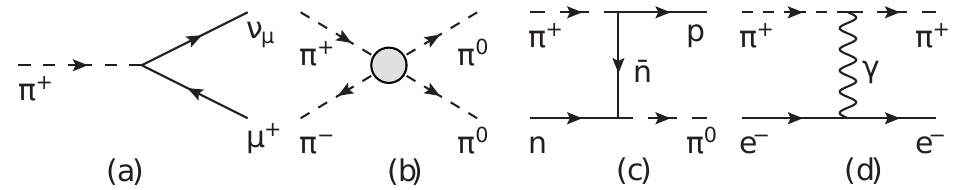}
    \caption{Diagrams of different interaction processes with metastable particles $Y$ in the primordial plasma: decay (a), annihilation $Y+\bar{Y}\to \text{SM}$ (b), the interaction with nucleons (c), and the elastic EM scattering that leads to the deposition of the $Y$'s kinetic energy in the EM plasma (d). Processes with the pion $\pi^{+}$ are considered as an example. The impact of the scattering off nucleons and annihilation process is demonstrated in Figs.~\ref{fig:probabilities-simple}-\ref{fig:rnu-HNL}.}
    \label{fig:interactions}
\end{figure*}

\begin{table*}[]
    \centering
    \begin{tabular}{|c|c|c|c|c|c|c|}
     \hline Particle & Decays & Annihilations & Nucleon interactions \\ \hline
       $\mu^\pm$ &  \makecell{$\tau = 2.2\cdot 10^{-6}\,\text{s}$ \\$e^\pm\overset{(-)}{\nu}_{e}\overset{(-)}{\nu}_{\mu}(100\%)$}  & \makecell{$\langle\sigma \beta\rangle\approx 6\cdot 10^{-2}\text{ GeV}^{-2}$ \\ $\gamma\gamma(25\%)$ \\ $e^{+}e^{-}(75\%)$} & \makecell{$\mu^- p\to n\nu_{\mu}$ \\ $\mu^+n\to p\bar\nu_\mu$\\ Subdominant} \\ \hline
       $\pi^{\pm}$ & \makecell{$\tau =2.6\cdot 10^{-8}\,\text{s}$\\ $\mu^\pm\overset{(-)}{\nu}_{\mu}: 100\%$}  & \makecell{$\langle\sigma \beta\rangle \approx 3-5\text{ GeV}^{-2}$\\ $2\pi^{0}( \approx 100\%)$} & \makecell{$\pi^{-}p\to nX$\\ $\langle\sigma \beta\rangle \approx 4-4.6\text{ GeV}^{-2}$ \\ $\pi^{+}n\to pX$\\ $\langle\sigma \beta\rangle \approx 4\text{ GeV}^{-2}$} \\ \hline
       $K^{\pm}$ & \makecell{$\tau = 1.23\cdot 10^{-8}\,\text{s}$ \\ $\mu\bar{\nu}_{\mu}(63\%)$ \\ $\pi^{0}l\bar{\nu}_{l}(8.4\%)$\\ $\pi^{+}\pi^{0}(20.7\%)$\\ $3\pi(7.4\%)$}  & \makecell{$\langle\sigma \beta\rangle \approx 44\text{ GeV}^{-2}$ \\ $\pi^{+}\pi^{-}(66.6\%)$ \\ $2\pi^{0}(33.3\%)$} & \makecell{$K^{-}p\to \mathcal{N} 2\pi$\\ $\langle\sigma \beta\rangle\approx 150\text{ GeV}^{-2}$\\ $K^{-}n\to \mathcal{N}2\pi$\\ $\langle\sigma \beta\rangle\simeq 10^{2}\text{ GeV}^{-2}$} \\ \hline
       $K_{L}$ & \makecell{$\tau =5.116\cdot 10^{-8}\,\text{s}$ \\ $\pi^{\pm} l^{\mp} \nu_{l}(67.6\%)$\\ $3\pi(30.6\%)$}  & Same as $K^{\pm}$ & \makecell{$K_{L}p\to \mathcal{N}2\pi$ \\ $\langle \sigma \beta\rangle\approx 42.5\text{ GeV}^{-2}$\\ $K_{L}n\to \mathcal{N}2\pi$ \\ $\langle\sigma \beta\rangle \approx 42.5\text{ GeV}^{-2}$} \\ \hline
       $K_{S}$ & \makecell{$\tau =0.89\cdot 10^{-10}\,\text{s}$\\ $2\pi^{0}(30.7\%)$\\ $\pi^{+}\pi^{-}(69.2\%)$}  & Same as $K^{\pm}$ & Same as $K_{L}$\\ \hline
    \end{tabular}
    \caption{Properties of the metastable particles in the primordial plasma. The meaning of the columns is as follows: the particle, its lifetime and decay modes, annihilation modes with their corresponding thermal-averaged cross-sections, and the same for the interactions with nucleons $\mathcal{N} = n,p$. For the thermal-averaged cross-sections, we provide the values at $T = 3\text{ MeV}$.}
    \label{tab:properties}
\end{table*}

A brief summary of the properties and interactions of the $Y$ particles is listed in table~\ref{tab:properties}, while the relevant interaction diagrams are shown in fig.~\ref{fig:interactions}.

We take the information about the decay modes from PDG~\cite{ParticleDataGroup:2022pth}, the interaction with nucleons from Ref.~\cite{Reno:1987qw}, and the results of this section for the annihilation channels.
Below, we describe them in detail for each of the particles.

\subsection{Muons}
\label{sec:muons}
The muon lifetime is $\tau_{\mu} \approx 1.2\cdot 10^{-6}\,\text{s}$ -- the largest among all the metastable particles. The only relevant decay mode is 
\begin{equation}
\mu^{-}\to e^{-}+\bar{\nu}_{e}+\nu_{\mu}.
\label{eq:decay-muon}
\end{equation}
The neutrino decay products may have energies as large as $E_{\nu} \approx m_{\mu}/2 \simeq 50\text{ MeV}$, which well exceeds the thermal neutrino energies at MeV temperatures.

The energy loss processes are 
\begin{equation}
\mu+\gamma\to \mu+\gamma \quad \text{and} \quad \mu+e\to \mu+e.
\end{equation}
The overall rate has the scaling 
\begin{equation}
\Gamma_{\text{loss}} = \langle \sigma_{\text{loss}}^{\mu} v\rangle n_{\text{EM}} \sim \frac{\alpha_{\text{EM}}^{2}}{m_{\mu}E_{\text{thermal}}}T_{\text{EM}}^{3},
\end{equation}
where $E_{\text{thermal}} \approx 3T$ is the mean energy of thermal particles. At $T = 1-5\text{ MeV}$, the rate is more than 9 orders of magnitude larger than the decay rate $\Gamma^{\mu}_{\text{decay}} = \hbar/\tau_{\mu}$. Because of this, we will assume that the muons are effectively at rest. The same conclusion holds for any other charged $Y$.

The annihilation processes are 
\begin{equation}
\mu^{+}+\mu^{-}\to e^{+}+e^{-} \quad \text{and} \quad \mu^{+}+\mu^{-}\to 2\gamma.
\end{equation}
They are thresholdless, and given that $m_{\mu}\gg m_{e}$, the thermal average $\langle\sigma \beta\rangle$ closely matches the zero-momentum cross-section:
\begin{equation}
    \langle\sigma^{\mu}_{\text{ann}} v\rangle \approx \sum_{i=ee,2\gamma}(\sigma^{\mu\to i}_{\text{ann}} v)_{i,p = 0} \approx  \frac{4\pi\alpha_{\text{EM}}^{2}}{m_{\mu}^{2}}.
\end{equation}

The annihilation rate $\Gamma_{\text{ann}}^{\mu} = \langle\sigma_{\text{ann}}^{\mu} v\rangle n_{\bar{\mu}}$ is also suppressed compared to the energy loss rate: the cross-section itself is smaller than energy loss one by the ratio $m_{\mu}/E_{\text{thermal}} \gg 1$, and the number density $n_{\bar{\mu}}$ of anti-muons (produced together with the muons) is much smaller than the thermal densities. This is because the instant $\bar{Y}$ number density, entering the annihilation rate $\Gamma_{\text{ann}} = n_{\bar{Y}}\langle\sigma v\rangle$, is principally bounded from above by what can be accumulated before decays. It is $n_{\bar{Y}} \lesssim n_{X}\frac{\tau_{Y}}{\tau_{X}}\ll n_{X}$ (see a discussion in Sec.~\ref{sec:simple})
, and hence is much smaller than the available $X$ number density (which is itself typically much smaller than the thermal number density). Note that generically, we assume that the same amounts of $Y$ and $\bar Y$ particles are produced by the $X$ decays, and the above argument on the annihilation rate applies equally to the charge-conjugated case.

Muons interact with nucleons $\mathcal{N} = n,p$ by 
\begin{equation}
\mu^-+p\to n+\nu_{\mu}, \quad
\mu^++n\to p+\bar\nu_{\mu}.
\end{equation}
They are mediated by the weak force, which, together with the tiny amount of nucleons, makes the processes irrelevant~\cite{Pospelov:2010cw,Fradette:2017sdd}.

\subsection{Charged pions}

The lifetime of the charged pion is $\tau_{\pi} = 2.6\cdot 10^{-8}\,\text{s}$, 2 orders of magnitude smaller than for the muon. The main decay mode is
\begin{equation}
\pi^{+}\to \mu^{+}+\nu_{\mu}
\label{eq:decay-pion}
\end{equation}
The neutrino produced by decays of the pion at rest has a monochromatic energy $E_{\nu} = (m_{\pi}^{2}-m_{\mu}^{2})/2m_{\pi} \approx 29.8\text{ MeV}$, which still greatly exceeds thermal neutrino energies.

The pion's energy loss rate is similar to the muon's one, being many orders of magnitude larger than the decay rate. As decaying pions inject muons, the evolution of $\pi$s and $\mu$s is coupled.

Despite the much smaller lifetime, the processes of the annihilation and the interaction with nucleons are important for the pions: the corresponding processes are driven by the strong force, which means a much larger cross-section. The dominant annihilation process is\footnote{The EM process, $\pi^{+}+\pi^{-}\to 2\gamma$, although being far from the kinematic threshold, is suppressed by two orders in magnitude.}
\begin{equation}
\pi^{+}+\pi^{-}\to 2\pi^{0} \,.
\end{equation}
It is close to the kinematic threshold, and the kinetic energy distribution of pions makes a non-negligible contribution to the cross-section. To compute it, we use the ChPT Lagrangian as implemented in~\cite{DallaValleGarcia:2023xhh}, and then average over thermally distributed pion energies using~\cite[eq.~(A.68)]{EscuderoAbenza:2020cmq}. Before averaging over energies, we find
\begin{equation}
    \sigma^{2\pi^{0}}_{\text{ann}} \beta = \frac{(10m_{\pi^{+}}^{2}+12p^{2}-m_{\pi^{0}}^{2})^{2}\sqrt{m_{\pi^{+}}^{2}-m_{\pi^{0}}^{2}+p^{2}}}{576\pi f_{\pi}^{4}(m_{\pi}^{2}+p^{2})^{\frac{3}{2}}},
\end{equation}
where $p$ is the momentum of the interacting pion in the center-of-mass frame and $f_{\pi}\approx 93\text{ MeV}$ is the pion decay constant. The thermal averaging increases the cross-section by a factor of 2 compared to the zero-momentum limit in the temperature range $T<5\text{ MeV}$. 

Let us now discuss interactions with nucleons. Since the pions are almost stopped, the most efficient processes are thresholdless. Those are~\cite{Reno:1987qw}
\begin{equation}
    \pi^{-}+p\to n+\pi^{0}/\gamma, \quad \pi^{+}+n\to p+\pi^{0}/\gamma \,.
    \label{eq:nucleon-interaction-pions}
\end{equation}
The thermal cross-sections behave as
\begin{align}
\langle\sigma^{\pi^{-}}_{p\to n} \beta\rangle \approx& \ 3.68\cdot F^{\pi}_{c}(T) \text{ GeV}^{-2}, \\ \langle\sigma^{\pi^{+}}_{n\to p} \beta\rangle \approx& \ 1.1\langle\sigma^{\pi^{-}}_{p\to n} v\rangle/F^{\pi}_{c}(T),
\label{eq:cross-section-nucleon-pions}
\end{align}
Here, 
\begin{equation}
F^{X}_{c}(T)= \frac{y_{X}}{1-\exp[-y_{X}]}, \quad y_{X} = 2\pi\alpha_{\text{EM}}/v_{\text{rel},pX},
\label{eq:sommerfeld}
\end{equation}
is the Sommerfeld enhancement, occurring because of the formation of a quasi-bound state of the oppositely charged $X$ and $p$ particles with the relative velocity $v_{\text{rel},pX} = | \bm{v}_{p}-\bm{v}_{X}|$.

The resulting $\langle \sigma^{\pi}_{p\leftrightarrow n} \beta\rangle$ is comparable to $\langle\sigma^{\pi}_{\text{ann}} \beta\rangle$.

\subsection{Kaons}
\label{sec:kaons}

The case of kaons is more complicated. There are four different kaons, $K^{\pm},K_{L},K_{S}$, with $K_{L/S}$ being admixtures of $K^{0}$ and $\bar{K}^{0}$. The lifetimes are $\tau_{K^{\pm}}\approx 1.23\cdot 10^{-8}\,\text{s}$, $\tau_{K_{L}}\approx 5.1\cdot 10^{-8}\,\text{s}$, and $\tau_{K_{S}}\approx 0.9\cdot 10^{-10}\,\text{s}$. All of them, except for $K_{S}$, have decay modes containing neutrinos. $K_{S}$ decays into a pair of pions; its lifetime is very small, and it does not have time to participate in any other interactions before decaying. The neutrino energy may be as large as $m_{K}/2$.

$K_{L}$s do not lose their kinetic energy before participating in any further interaction. Here and below, we will treat them as particles-at-rest for simplicity. We argue this in the following way. If including the finite energy distribution of kaons, the decay probability decreases with the $\gamma$ factor (due to time dilation). On the other hand, the probabilities of the other processes would generically increase, as we enlarge the available scattering phase space. Therefore, our approximation would overestimate the decay probability of $K_{L}$. However, as we study GeV-scale LLPs, the impact of these changes would not be significant, which justifies the approach.

The dominant kaon annihilation processes are 
\begin{equation}
K^{+}+K^{-}\to \pi^{+}+\pi^{-}, \quad K^{+}+K^{-}\to 2\pi^{0}
\end{equation}
(and the same for $K_{L}, K_{S}$ particles). Since $m_{K}-m_{\pi}\gg 3T$, the reactions are far from threshold, and we may safely approximate their cross-sections $\sigma v$ by the zero-momentum result:
\begin{equation}
    \langle\sigma^{K}_{\text{ann}} \beta\rangle \approx \frac{\sqrt{m_{K}^{2}-m_{\pi}^{2}}(m_{d}(10m_{K}^{2}+m_{\pi}^{2})+m_{\pi}^{2}m_{s})^{2}}{3072\pi f_{\pi}^{4}m_{d}^{2}m_{K}^{3}},
\end{equation}
with the numeric value $\approx 44\text{ GeV}^{-2}$. It is a factor of 10 larger than $\langle \sigma^{\pi}_{\text{ann}} \beta\rangle$, because the reaction is far from the threshold.

The interaction processes with nucleons $\mathcal{N}$ are much more complicated than in the pion case. The thresholdless processes exist only for $K_{L},K^{-}$, and go via the intermediate $\Lambda/\Sigma$ resonances~\cite{Reno:1987qw,Pospelov:2010cw}:
\begin{align}
    K^{-} + \mathcal{N} \to& \ \Lambda/\Sigma + \pi \to \mathcal{N}' + 2\pi, \\ \quad K_{L}+\mathcal{N} \to& \ \Lambda/\Sigma +\pi \to \mathcal{N}'+2\pi .
    \label{eq:nucleon-interaction-kaons}
\end{align}
The absence of such processes for $K^{+}$ follows from the fact that they would require resonances with positive baryon number and strangeness, that do not exist. The asymmetry in the evolution of $K^{+},K^{-}$ induces an asymmetry in the energy distributions of neutrinos and antineutrinos; we will return to this phenomenon in Sec.~\ref{sec:asymmetry}.

The thermal cross-sections (here assuming that $K_{L}$ is at rest) are~\cite{Reno:1987qw}
\begin{alignat}{3}
    \langle{\sigma^{K^-}_{p\rightarrow n}\beta}\rangle & \approx 79 F^K_c(T) \text{ GeV}^{-2}, \quad & \langle{\sigma^{K^-}_{n\rightarrow p}\beta}\rangle & \approx 66 \text{ GeV}^{-2}, \\
    \langle{\sigma^{K^-}_{p\rightarrow p}\beta}\rangle & \approx 37 F^K_c(T) \text{ GeV}^{-2}, & \langle{\sigma^{K^-}_{n\rightarrow n}\beta}\rangle & \approx 88 \text{ GeV}^{-2}, \\
    \langle{\sigma^{K_L}_{p\rightarrow n}\beta}\rangle & \approx 18 \text{ GeV}^{-2}, & \langle{\sigma^{K_L}_{n\rightarrow p}\beta}\rangle & \approx 18\text{ GeV}^{-2} .
\end{alignat}
Here, $F_{c}$ is given by eq.~\eqref{eq:sommerfeld}. 

Kaon decays, annihilations, and interaction with nucleons inject charged pions and/or muons, which do not transfer all their energy to the EM plasma. Therefore, the evolution of $K$, $\mu$, and $\pi$ populations is coupled.

\section{Two-step approach}
\label{sec:approach}

In this section, we discuss our approach to studying the evolution of the metastable particles $Y$ in the primordial plasma and their impact on neutrinos. 

We assume a generic scenario when $Y$\!s are injected by decays of some hypothetical LLP, denoted by $X$, at MeV temperatures. We are agnostic about the origin of $X$ and parameterize its number density as 
\begin{equation}
n_{X} = n_{X,0}\left(\frac{a(t_{0})}{a(t)}\right)^{3}\exp\left[-\frac{t-t_{0}}{\tau_{X}}\right]
\label{eq:nX}
\end{equation}
Here, $n_{X,0}$ is the number density at some initial time $t_{0}$, $a(t)$ is the scale factor of the Universe, and $\tau_{X}$ is its lifetime.

Decays into $Y$\!s are only possible if $m_{X}>m_{Y}\gg 3T$. This means that the LLPs we consider have to be out-of-equilibrium at the temperatures of interest; otherwise, their abundance would be exponentially suppressed. As for the LLP lifetimes, our main interest is in the range $\mathcal{O}(0.01-10)\,\text{s}$. On the one hand, it covers the temperatures from the beginning of the neutrino decoupling to shortly after (in $\Lambda$CDM). On the other hand, this is also the temperature range where the metastable particles may prefer to disappear without decaying.

In general, $n_{X,0}$ is an independent parameter, but for particular models with only two parameters -- mass $m_{X}$ and $\tau_{X}$ -- it may be uniquely fixed: $n_{X,0} = n_{X,0}(m_{X},\tau_{X})$. In the rest of the paper, we will explore both of these scenarios to cover as broad a range of models as possible.

In order to study the dynamics of the metastable particles and neutrinos, we follow a \textit{two-step approach}:
\begin{itemize}
\item[1.] We trace the evolution of $Y$ particles in the expanding Universe, utilizing a simplified description of the neutrino dynamics from~\cite{EscuderoAbenza:2020cmq} as seed; details are described in Sec.~\ref{sec:metastable-dynamics} (see also Appendix~\ref{app:integrated}). 
\item[2.] We include the calculated evolution of $Y$\!s from step 1 in the form of time-dependent decay probabilities in the source term of the solver of the neutrino Boltzmann equation in the momentum space. Then, we carefully trace the evolution of neutrinos and the expansion of the Universe; see sec.~\ref{sec:unintegrated} and appendix~\ref{app:unintegrated}. 
\end{itemize}

This factorization is meaningful because the evolution of $Y$\!s is weakly affected by details of the equilibration between neutrinos and EM plasma: it is mainly sensitive to the scale factor, which is determined by the overall energy density of the Universe.

\section{Step I. Dynamics of metastable particles}
\label{sec:metastable-dynamics}

\subsection{System of equations}

Let us now construct the system of equations for the $Y$ abundances. Most of the $Y$\!s are charged and, therefore, effectively at rest; given this, it is adequate to consider the system of coupled integrated Boltzmann equations on their number densities.

As we have discussed in Sec.~\ref{sec:Y}, we have to solve the system for all $Y$\!s simultaneously, given that their dynamics are coupled: heavier $Y$\!s produce lighter ones because of decay, annihilation, or interactions with nucleons. Then, the resulting equations for the given $Y$ and its antiparticle $\bar{Y}$ take the form
\begin{widetext}
\begin{equation}
\begin{cases} \frac{dn_{Y}}{dt}+3Hn_{Y} = \frac{n_{X}}{\tau_{X}}N^{X}_{Y} - \frac{n_{Y}}{\tau_{Y}}-n_{Y}n_{\bar{Y}}\langle \sigma_{\text{ann}}^{Y} v\rangle+\left(\frac{dn_{Y}}{dt}\right)_{\mathcal{N}}+\sum_{Y'\neq Y}n_{Y'}\Gamma_{Y'\to Y}, \\ \frac{dn_{\bar{Y}}}{dt}+3Hn_{\bar{Y}} = \frac{n_{X}}{\tau_{X}}N^{X}_{\bar{Y}} - \frac{n_{\bar{Y}}}{\tau_{Y}}-n_{\bar{Y}}n_{Y}\langle \sigma_{\text{ann}}^{Y} v\rangle+\left(\frac{dn_{\bar{Y}}}{dt}\right)_{\mathcal{N}}+\sum_{Y'\neq Y}n_{Y'}\Gamma_{Y'\to \bar{Y}}.
        \end{cases}
        \label{eq:system-Y}
        \end{equation}
\end{widetext}
The meaning of the terms is as follows. 
\begin{itemize}
\item The second term on the left-hand side appears due to the expansion of the Universe. $H=\sqrt{8\pi\rho/(3m_{\rm Pl})}$ is the Hubble parameter, with the Planck mass $m_{\rm Pl}=1.2\times 10^{19}\ {\rm GeV}$ and the energy density of the Universe $\rho$.
\item $\frac{n_{X}}{\tau_{X}}N^{X}_{Y}$ is the injection from decays of $X$. Apart from direct decays, we also include secondary contributions $X \to Z \to Y$, where $Z$\! are ultra short-lived particles with $\tau_{Z} \ll 10^{-8}\,\text{s}$: $K_{S},\rho^{0},\eta,\omega$, etc. $N^{X}_{Y}$  is the amount of $Y$\!s per $X$ decay:
\begin{equation}
N^{X}_{Y} = \sum_{i}\text{Br}_{i}\cdot N_{Y}^{i},
\label{eq:NY}
\end{equation}
with $\text{Br}_{i}$ being the branching ratio of the given decay channel $i$, and $N^i_Y$ denoting the number of $Y$\!s produced per this channel.
\item The 2nd and 3rd terms on the r.h.s.\ of eq.~\eqref{eq:system-Y} describe direct decays and annihilations of $Y$, respectively. 
\item
$\left(\frac{dn_{Y/\bar{Y}}}{dt}\right)_{\mathcal{N}}$ is the evolution due to the interaction with nucleons $\mathcal{N} = p,n$:
\begin{equation}
\left(\frac{dn_{Y/\bar{Y}}}{dt}\right)_{\mathcal{N}} = -n_{Y/\bar{Y}}\sum_{\mathcal{N}}n_{\mathcal{N}}\langle \sigma^{Y/\bar{Y}}_{\mathcal{N}}v\rangle .
\end{equation}
The interaction processes include the $p\leftrightarrow n$ conversion as well as the processes that do not change the $\mathcal{N}$ type.
\item
The summand $\sum_{Y'\neq Y}n_{Y'}\Gamma_{Y'\to Y}$ takes into account decay, annihilation, and nucleon interaction processes involving the metastable particles $Y'\neq Y$ with $m_{Y'}>m_{Y}$:
\begin{multline}
\Gamma_{Y'\to Y} = \frac{1}{\tau_{Y'}}N^{Y',\text{decay}}_{Y}+\\+n_{\bar{Y'}}\langle\sigma_{\text{ann}}^{Y'} v\rangle N^{Y',\text{ann}}_{Y} + \sum_{\mathcal{N}}n_{\mathcal{N}}\langle\sigma^{Y'}_{\mathcal{N}} v\rangle N^{Y',\mathcal{N}}_{Y},
\end{multline}
with $N^{Y',\text{decay}}_{Y},N^{Y',\text{ann}}_{Y},N^{Y',\mathcal{N}}_{Y}$ being the amounts of $Y$ produced per given process. We calculate them using~\cite{ParticleDataGroup:2022pth} for decays, \cite{Reno:1987qw} for the interaction with nucleons, and this work for the annihilation.
\end{itemize}

The system~\eqref{eq:system-Y} has to be supplemented by the equations governing the evolution of EM and neutrino populations, the scale factor, and the nucleon number densities. The first two we calculate using the method from~\cite{Escudero:2018mvt,EscuderoAbenza:2020cmq}. It assumes that throughout their evolution, the neutrinos $\nu_{\alpha}$ always have Fermi-Dirac shape of the energy spectrum, parametrized by a time-dependent temperature $T_{\nu_{\alpha}}(t)$. Under this simplification, it is possible to integrate the neutrino Boltzmann equation in the momentum space and get the system of equations on the neutrino and EM temperatures $T_{\nu_{e}},T_{\nu_{\mu}},T_{\nu_{\tau}},T$ and scale factor $a$ in the presence of decaying LLPs. Throughout the text, we call the approach by the \textit{integrated} method to solve neutrino Boltzmann equations, as opposed to the unintegrated method that we will consider in sec.~\ref{sec:unintegrated} (see Appendix~\ref{app:integrated} discussing our implementation of the integrated approach, including incorporating neutrino oscillations). When calculating the source terms for neutrinos and the EM particles within this method, we assume that all $Y$\!'s energy goes to the EM plasma.\footnote{By varying this assumption and checking how the resulting $Y$ dynamics change, we have explicitly verified that this assumption is not important.}

Knowing the resulting dynamics of the scale factor $a$ determines the evolution of the $X$'s number density~\eqref{eq:nX} and the baryon-to-photon ratio
\begin{equation}
    \eta_{B}(T) = \eta_{B,\text{Planck}}\cdot \left(\frac{a(T_{\text{CMB}})T_{\text{CMB}}}{aT}\right)^{3},
\end{equation}
where $\eta_{B,\text{Planck}} = 7.06\cdot 10^{-10}$ is fixed by the CMB measurements performed with Planck~\cite{Planck:2018vyg}.

For the nucleon number density, we start with the definition
\begin{equation}
n_{\mathcal{N}}(t) \equiv n_{B}(t)X_{\mathcal{N}}(t) = n_{\gamma}\eta_{B}(t)\cdot X_{\mathcal{N}}(t),
\end{equation}
where $n_{B}$ is the baryon number density, and $X_{\mathcal{N}} \equiv n_{\mathcal{N}}/n_{B}$ is the relative fraction of the given nucleon ($X_{n}+X_{p} = 1$). The latter obeys the equation
\begin{multline}
\frac{dX_{n}}{dt} = -X_{n}\left(\Gamma^{\nu,e}_{n\to p}+\sum_{y = Y,\bar{Y}}n_{y}\langle\sigma^{y}_{n\to p} v\rangle\right)+\\ + (1-X_{n})\left(\Gamma^{\nu,e}_{p\to n}+\sum_{y = Y,\bar{Y}}n_{y}\langle\sigma^{y}_{p\to n} v\rangle\right),
\label{eq:Xn-evolution}
\end{multline}
where $\Gamma^{\nu,e}_{n\leftrightarrow p}$ are rates of the weak conversion processes with neutrinos and electrons, while $n_{Y}\langle\sigma_{p\leftrightarrow n}^{Y} v\rangle$ are those driven by the $Y$ particle. The latter processes are part of the total nucleon interaction rates $\langle\sigma_{\mathcal{N}}^{Y} v\rangle$:
\begin{equation}
\langle\sigma_{\mathcal{N}}^{Y} v\rangle = \langle\sigma_{\mathcal{N}\to \mathcal{N}}^{Y} v\rangle  +\langle\sigma_{\mathcal{N}\to \mathcal{N}'}^{Y} v\rangle.  
\end{equation}
If $Y$ is a meson, it completely dominates the evolution of $X_{n}$ until the instant $Y$ population is suppressed by many orders of magnitude compared to the neutrino number density~\cite{Boyarsky:2020dzc}. This is because of two factors. First, the meson-driven conversion cross-section is 16 orders of magnitude larger than the cross-section of the weak conversion. Second, at MeV temperatures, the probability of $Y$ interactions with nucleons is comparable with its decay probability, so there is no a priori suppression. Therefore, in practice, the weak $p\leftrightarrow n$ conversion rates may be dropped from eq.~\eqref{eq:Xn-evolution}.

The solution for $X_{n}$ may be obtained by setting the right-hand-side of eq.~\eqref{eq:Xn-evolution} to zero (the so-called dynamic equilibrium):\footnote{We have validated the dynamical equilibrium solutions for $X_{n}$ and $n_{Y}$ (eq.~\eqref{eq:Yabundance}) by computing first the exact solutions and comparing them with the approximate solution given by the dynamic equilibrium.}
\begin{equation}
    X_{n} \approx \frac{\sum_{y}n_{y}\langle\sigma_{p\to n}^{y} v\rangle}{\sum_{y}n_{y}\langle\sigma_{p\to n}^{y} v\rangle+\sum_{y}n_{y}\langle\sigma_{n\to p}^{y} v\rangle}.
    \label{eq:Xn-dynamic-equilibrium}
\end{equation}
Once we solve the coupled system of equations for $\mu,\pi,K,X_{n}$, we may compute time-dependent probabilities to decay and disappear by annihilating or interacting with nucleons:
\begin{align}
    P^{Y}_{\text{decay}}(t) &= \ \frac{\tau_{Y}^{-1}}{\tau_{Y}^{-1}+\sum_{\mathcal{N}}n_{\mathcal{N}}\langle \sigma^{Y}_{\mathcal{N}} v\rangle+n_{\bar{Y}}\langle \sigma_{\text{ann}}^{Y}v\rangle}, \label{eq:Pdec} \\ 
    P^{Y}_{\text{ann}}(t) &= \ \frac{n_{\bar{Y}}\langle \sigma_{\text{ann}}^{Y}v\rangle}{\tau_{Y}^{-1}+\sum_{\mathcal{N}}n_{\mathcal{N}}\langle \sigma^{Y}_{\mathcal{N}} v\rangle+n_{\bar{Y}}\langle \sigma_{\text{ann}}^{Y}v\rangle}, \label{eq:Pann}\\
    P^{Y}_{\mathcal{N}}(t) &= \ \frac{\sum_{\mathcal{N}}n_{\mathcal{N}}\langle \sigma^{Y}_{\mathcal{N}} v\rangle}{\tau_{Y}^{-1}+\sum_{\mathcal{N}}n_{\mathcal{N}}\langle \sigma^{Y}_{\mathcal{N}} v\rangle+n_{\bar{Y}}\langle \sigma_{\text{ann}}^{Y}v\rangle} \label{eq:Pnucl}
\end{align}
These probabilities serve as an input to calculate the impact on the neutrino and EM populations of the primordial plasma. We separate annihilations and interactions with nucleons, as the latter are very important for studying the impact of $Y$ on BBN.

Assuming that we have computed the decay probability $P^{Y}_{\text{decay}}(t)$, the number density of $Y$\!s available for decays is again given by the dynamical equilibrium:
\begin{equation}
n_Y(t) = n_{X}(t)N_{Y}^{X}\frac{\tau_Y}{\tau_{X}}P_{\rm decay}^{Y}(t)
\label{eq:Yabundance}
\end{equation}

We provide the implementation of this system and its solution for generic LLPs in a \texttt{Mathematica} code.\footnote{Available on \faGithub \cite{GitHub-metastable} and Zenodo~\cite{Zenodo}.} Details on the code may be found in Appendix~\ref{app:code} and on the GitHub repository page.

\subsection{Simple estimates of $Y$ evolution}
\label{sec:simple}

Let us make a simplified analysis that allows us to understand the impact of annihilation and interaction with nucleons. 
First, let us neglect the influence of $X$ particles on the Hubble expansion rate. Then, we may use the standard formula $a(t)\propto \sqrt{t}$ and $H(t) =\dot{a}/a  = 1/2t$ for the radiation-dominated Universe, as well as the standard cosmological value for the baryon-to-photon ratio $\eta_{B}(1\text{ MeV})\approx 1.7\cdot 10^{-9}$. Next, let us assume that various $Y$\!s evolve independently from each other. With all these approximations, we can still qualitatively describe the dynamics of the populations of $Y$ and its antiparticle $\bar{Y}$, while presenting results in a simple form.

Similarly to the case of $X_{n}$, we may solve the system~\eqref{eq:system-Y} analytically in the regime of dynamic equilibrium, when all the processes are much faster than the Hubble expansion.\footnote{Note that the form of the expression~\eqref{eq:analytic-solution} differs from~\eqref{eq:Yabundance}. This is because in~\eqref{eq:Yabundance} we assume that the decay probability $P^{Y}_{\text{decay}}$ has been previously computed numerically. The latter includes $n_{\bar{Y}}$, which is tightly related to $n_{Y}$.} Assuming $n_{Y} = n_{\bar{Y}}$, we get
\begin{multline}
    n_{Y} = \frac{\sqrt{\frac{4n_{X}\langle\sigma^{Y}_{\text{ann}} v\rangle}{\tau_{X}}+\left(\Gamma_\mathcal{N} +\tau^{-1}_{Y}\right)^{2}} - \Gamma_\mathcal{N} -\tau^{-1}_{Y}}{2\langle\sigma_{\text{ann}}^{Y} v\rangle}  \,,
    \label{eq:analytic-solution}
\end{multline}
where we have defined an effective interaction rate with nucleons as
\begin{equation}
    \Gamma_\mathcal{N} \equiv \sum_{\mathcal{N}}n_{\mathcal{N}}\langle \sigma_{\mathcal{N}}^{Y} v\rangle
\end{equation}

Now, let us analyze this solution by considering two limiting cases: $n_{Y}\langle\sigma^{Y}_{\text{ann}} v\rangle \gg \Gamma_\mathcal{N}$, meaning that annihilations dominate over the interactions with nucleons, and $n_{Y}\langle\sigma^{Y}_{\text{ann}} v\rangle \ll \Gamma_\mathcal{N}$, which is the opposite. 

For the first case, we can estimate the relative importance of decays and annihilations by considering
\begin{equation}
   n_{Y} =\frac{1}{2\langle\sigma^{Y}_{\text{ann}} v\rangle \tau_Y}
   \left[\sqrt{\frac{4}{\epsilon_\text{ann}}+1} - 1\right],
\end{equation}
where we have used eq.~\eqref{eq:analytic-solution} in the limit $\Gamma_\mathcal{N} = 0$ and defined\footnote{$\epsilon_{\text{ann}}$ may be understood in the following way. Consider an instant injection of $Y$ from $n_{X}$ during time $\sim \tau_{Y}$; during this period, decays do not deplete the $Y$ population. Then, let us assume a priori that the annihilation does not prevent accumulating $\bar{Y}$ during this time, so $n_{\bar{Y}}\approx \frac{n_{X}}{\tau_{X}}\tau_{Y}$. For the ratio of $\Gamma_{\text{decay}}$ and $\Gamma_{\text{ann}} = n_{\bar{Y}}\langle \sigma v\rangle_{\text{ann}}$, one then gets eq.~\eqref{eq:ratio1}.}
\begin{equation}
\epsilon_{\text{ann}} = \frac{\tau_{Y}^{-2}}{\frac{n_{X}}{\tau_{X}}\langle \sigma^{Y}_{\text{ann}} v\rangle}
\label{eq:ratio1}
\end{equation}
For the second case ($n_{Y}\langle\sigma^{Y}_{\text{ann}} v\rangle \ll \Gamma_\mathcal{N}$), we can directly compare the decay rate to the rate of the interaction with nucleons, which are both independent of the abundance $n_Y$:
\begin{equation}
\epsilon_{\mathcal{N}} = \frac{\tau_{Y}^{-1}}{\Gamma_\mathcal{N}}
\label{eq:ratio2}
\end{equation}
Hence, in both cases, a small value for the ratios \eqref{eq:ratio1} and \eqref{eq:ratio2} implies that $Y$ decays are much less efficient than the competing processes (annihilations or interactions with nucleons, respectively).

\begin{table}[]
    \centering
    \begin{tabular}{|c|c|c|c|c|c|c|}
       \hline Particle & $\epsilon_{\text{ann}}$ & $ \epsilon_{\mathcal{N}}$  \\ \hline
      $\mu^{\pm}$ & $3.4\cdot 10^{-4}$ &  $\gg 1$ \\ \hline
      $\pi^{\pm}$ & $4.1\cdot 10^{-2}$ & $1.15$\\ \hline
      $K^{-}$ & $1.4\cdot 10^{-2}$ & $3.4\cdot 10^{-2}$ \\ \hline
      $K^{+}$ & $1.4\cdot 10^{-2}$ & $\gg 1$ \\ \hline
      $K_{L}$ & $8.6\cdot 10^{-4}$ & $6.8\cdot 10^{-2}$ \\ \hline
      $K_{S}$ & $2.8\cdot 10^{2}$ & $40.$  \\ \hline
    \end{tabular}
    \caption{Ratios~\eqref{eq:ratio1},~\eqref{eq:ratio2} for $T = 3\text{ MeV}$, $\tau_{X} = 0.05\,\text{s}$, and the LLP number density given by eq.~\eqref{eq:setup-qualitative}. The cross-sections are taken from Sec.~\ref{sec:Y}. We assumed for simplicity $n_{p}(T) \approx n_{B}/2\approx \eta_{B}(T)\cdot n_{\gamma}/2$, with $\eta_{B}(T\gg m_{e})\approx 1.7\cdot 10^{-9}$.}
    \label{tab:rates-ratios}
\end{table}
Let us consider the reference choice
\begin{equation}
n_{X,0} = 0.1\cdot n_{\text{UR}}(T_{0})= 0.1\cdot\frac{\zeta(3)}{\pi^{2}}T_{0}^{3},
\label{eq:setup-qualitative}
\end{equation}
where $n_{\text{UR}}$ is the number density of a scalarultrarelativistic particle in equilibrium at the given temperature, 
and $\tau_{X} = 0.03\,\text{s}$. The values of the quantities~\eqref{eq:ratio1} and~\eqref{eq:ratio2} are shown in Table~\ref{tab:rates-ratios}.
They clearly imply that the dynamics of stopped pions, $K^{\pm}$, muons, and $K_{L}$ may be driven not by decays but by annihilations or interactions with nucleons. For example, the smallness of $\epsilon_{\text{ann}}$ suggests that the particle prefers to annihilate rather than decay. The exceptions are short-lived $K_{S}$: their tiny lifetime allows them to decay before interacting. 

\begin{figure}[t!]
    \centering
    \includegraphics[width=0.45\textwidth]{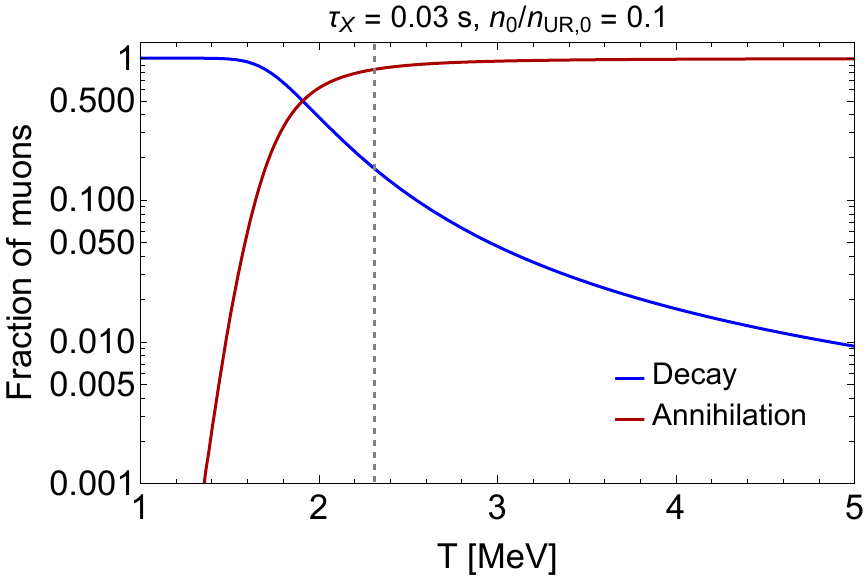} \\ \includegraphics[width=0.45\textwidth]{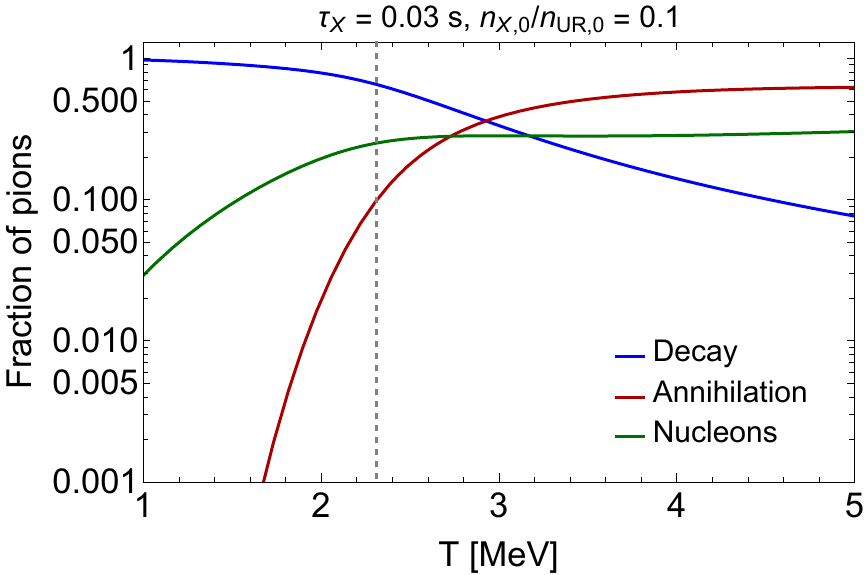}
    \caption{The yields of muons (top) and pions (bottom) that would decay, annihilate, or interact with the nucleons (Eqs.~\eqref{eq:Pdec}-\eqref{eq:Pnucl}) if injected by a decaying particle $X$ with the number density given by eq.~\eqref{eq:nX} and the lifetime $\tau_{X} = 0.03\,\text{s}$. The results are obtained using the simplified consideration presented in Sec.~\ref{sec:simple}, in order to be easily reproducible. The vertical dashed line shows the moment of time at which the comoving density of $X$ becomes $0.01$ of the $n_{X,0}$, such that the dynamics of $X$ and its decay products already do not affect the Universe.}
    \label{fig:probabilities-simple}
\end{figure}

The impact of the scattering processes significantly depends on the number density of the interacting counterparts -- $\bar{Y}$ for annihilation and baryons for the nucleon interactions. Both $n_{\bar{Y}}$ and $n_{B}$ are suppressed at low temperatures as $a^{-3}\sim T^{3}$. In addition, the $\bar{Y}$ number density, entering the annihilation rate for $Y$\!s, gets exponentially suppressed at times $t\gg \tau_{X}$, so the drop in $P^{Y}_{\rm ann}$ would be much faster than in $P_{\mathcal{N}}^{Y}$. To account for these effects, we will use the solution~\eqref{eq:analytic-solution} and obtain the probabilities~\eqref{eq:Pdec}-\eqref{eq:Pnucl} for muons and pions. They are shown in fig.~\ref{fig:probabilities-simple} for the setup~\eqref{eq:setup-qualitative}. 
For the particular parameters, decays are strongly suppressed at high temperatures but become dominant at a temperature determined by the properties of $X$. A higher $X$ number density is associated with a lowering of this temperature.

\section{Step II. Evolution of neutrinos}
\label{sec:unintegrated}

To trace the evolution of neutrinos and EM plasma in the presence of decaying relics, we need to solve the system of equations governing the evolution of the neutrino distribution function and the EM plasma temperature. This system must incorporate the dynamics of the metastable decay products discussed in this paper. 

The most accurate way to follow the evolution of out-of-equilibrium neutrinos, including neutrino oscillations, would be solving the evolution equations for the 3-state density matrix of neutrinos, called quantum kinetic equations (QKEs)~\cite{Vlasenko:2013fja,Volpe:2013uxl,Serreau:2014cfa,Akita:2020szl}. However, the accuracy of that approach is far better than needed, given the expected uncertainty of future CMB observations, and it is also computationally expensive.

At MeV temperatures, the rates of neutrino oscillations are much faster than the weak interaction rates. Because of this, one can separate the weak interactions from the oscillations and describe the neutrino evolution as quasi-classical unintegrated Boltzmann equations~\cite{Hannestad:1995rs,Dolgov:2002wy} on the neutrino distribution function $f_{\nu_{\alpha}}(p,t)$:
\begin{equation}
\frac{\partial f_{\nu_{\alpha}}}{\partial t}-Hp\frac{\partial f_{\nu_{\alpha}}}{\partial p}=\sum_{\beta=e,\mu,\tau}\langle P_{\alpha\beta}\rangle I_\beta[p,f_{\nu_{\alpha}}].
\label{eq:BE}
\end{equation}
Here, $\langle P_{\alpha\beta}\rangle (p,T)$ is the time-averaged oscillation probability depending on the temperature of the EM plasma $T$ and neutrino energy, accounting for the matter corrections to the mixing angles, expressed in terms of neutrino energy and EM plasma temperature (see Sec. 3 in ref.~\cite{Sabti:2020yrt} and ref.~\cite{Strumia:2006db}). $I_\alpha$ is the collision term, which describes the details of scattering and annihilation of $\nu_\alpha$, as well as the decays of LLPs:
\begin{multline}
I_\alpha = \frac{1}{2E_{\nu_\alpha}}\sum \int \prod_{i=2} \frac{d^3 p_i}{(2\pi)^3 2E_i} \prod_{f=1} \frac{d^3 p_f}{(2\pi)^3 2E_f} \\ 
\times S |\mathcal{M}|^2F[f](2\pi)^4\delta^{(4)} \left(\sum_{i=1} p_i- \sum_{f=1} p_f \right).
\label{eq:Collterm}
\end{multline}
The first sum runs over all possible interaction processes, including $\nu_\alpha$ as $i=1$. The integral is performed for all possible states for $\nu_\alpha$ with momentum $p_1$. $i,f$ denote the initial and final states for a process. $S|\mathcal{M}|^2$ is the corresponding squared matrix element times the symmetry factor $S$.\footnote{See table 3 in ref.~\cite{Sabti:2020yrt} for the specific formula of $S|\mathcal{M}|^2$ for relevant processes of neutrinos at MeV temperature.} $F[f]$ is the quantum statistical factor to describe the population of the medium:
\begin{equation}
F[f]=\prod_{i=1}(1\mp f_i)\prod_{f=1} f_f - \prod_{i=1} f_i \prod_{f=1} (1\mp f_f),
\end{equation}
where $f_{i,f}$ are the momentum distribution for the $i,f$-th particle. $(1-f)$ is the Pauli-blocking factor for fermions while $(1+f)$ is the Bose-enhancement factor for bosons.

To close the equations for the system of the plasma, it is necessary to know the evolution equation of the electromagnetic plasma (i.e., the EM temperature). It is described by the continuity equation (the energy conservation law), including the energy densities of the SM plasma and the decaying LLPs:
\begin{equation}
\frac{d\rho}{dt}=-3H(\rho+P),
\label{eq:CE}
\end{equation}
where $\rho$ and $P$ are the total energy density and pressure for the plasma, respectively. 
As in ref.~\cite{Akita:2020szl}, we include thermal QED corrections in eq.~\eqref{eq:CE} following~\cite{Heckler:1994tv,Fornengo:1997wa,Kapusta:2006pm,Bennett:2019ewm}. Further discussion on our implementation, running time, and limitations is given in appendix~\ref{app:unintegrated}.

Our Boltzmann solver has been extensively tested against a completely independent method to calculate the evolution of neutrino distribution functions, namely the neutrino Direct Simulation Monte-Carlo method developed in Refs.~\cite{Ovchynnikov:2024rfu,Ovchynnikov:2024xyd}. Examples of the cross-checks include the evolution of the neutrino spectrum under the initial conditions of different neutrino and EM plasma temperatures, and injections of high-energy neutrinos. The agreement in the evolution of characteristic quantities, such as the energy densities, is at the level of $0.1\%$. Moreover, the unintegrated quantities, such as neutrino energy spectra, also agree quite well. 

\subsection{Simple case study: neutrino distortions matter}

\begin{figure}[t!]
    \centering
    \includegraphics[width=\linewidth]{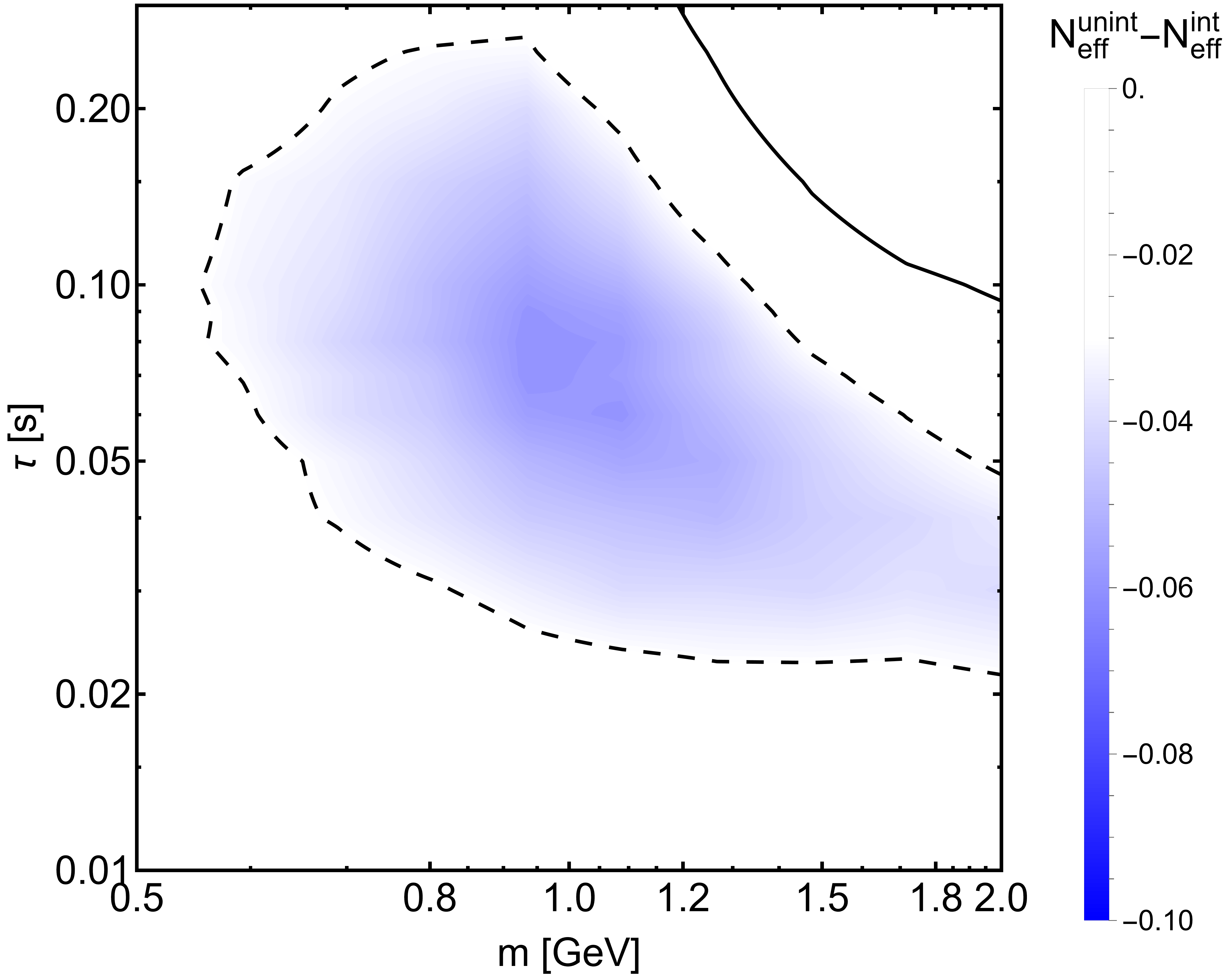}
    \caption{The behavior of  $N_{\text{eff}}$ under the scenario with an LLP decaying solely into the EM particles (see text for details). The plot shows the behavior of the deviation $N_{\text{eff}}^{\text{unint}} -N_{\text{eff}}^{\text{int}}$, where $ N_{\text{eff}}^{\text{unint}}$ has been obtained using the unintegrated method to solve the neutrino Boltzmann equation described in Sec.~\ref{sec:unintegrated}, whereas $\Delta N_{\text{eff}}^{\text{int}}$ is calculated utilizing the integrated approach from~\cite{EscuderoAbenza:2020cmq} including neutrino oscillations (see~\ref{app:integrated}).  The solid black line indicates zero deviation, whereas the dashed black line shows $N_{\text{eff}}^{\text{unint}} -N_{\text{eff}}^{\text{int}} = -0.03$.}
    \label{fig:cross-check-1}
\end{figure}

Using the momentum-dependent solver is necessary even in the simplest scenarios with no high-energy neutrino injections. To illustrate this point, we consider a relic $X$ decaying solely into electromagnetic particles as an example (an example of such an LLP is a light Higgs-like scalar or axion-like particle). We will compare the predictions of our code with the integrated approach to solve the neutrino Boltzmann equations we introduced in the previous section~\ref{app:integrated}.

Fig.~\ref{fig:cross-check-1} shows the quantity $N_{\text{eff}}^{\text{unint}}-N_{\text{eff}}^{\text{int}}$, i.e., the difference between the values of $N_{\text{eff}}$ predicted by the unintegrated and integrated approaches, considering different mass and lifetime of the LLP. As for the initial abundance of the LLP, we use 
\begin{equation}
n_{\text{LLP}}(T = 20\text{ MeV}) = 0.05\cdot n_{\text{UR}}(20\text{ MeV})
\end{equation}
Naively, the purely EM decays do not disturb the thermal shape of the neutrino spectrum, and then the predictions of the unintegrated and integrated methods should match. We confirm this expectation in the lifetimes ranges $\tau\lesssim 0.02\text{ s}$ (where there is no impact on $N_{\text{eff}}$ at all) and $\tau \gtrsim 0.2\text{ s}$. 

However, there is a sizeable deviation in the intermediate domain. Its origin is accumulating distortions in the neutrino sector during the equilibration with the EM plasma (see also the discussion in ref.~\cite{Ovchynnikov:2024rfu}). Namely, the neutrino-EM interaction rate is a growing function of the neutrino energy, so high-energy neutrinos equilibrate better. Distortions are mainly accumulated in the temperature range $2\text{ MeV}\lesssim T\lesssim 5\text{ MeV}$ (translating to the lifetime range mentioned above) when neutrinos are already partially decoupled but still interact efficiently with the EM particles. They destroy the main approximation of the integrated approach, which leads to the discrepancy (see also \cite{Sabti:2020yrt}).

The magnitude of the deviations between the unintegrated and integrated methods may go beyond the error bars in the determination of $N_{\text{eff}}$ by future CMB observations, such as Simons Observatory. Therefore, even such relatively simple scenarios with LLPs require accurate numeric studies of how the neutrino population is affected.

\section{Qualitative impact of metastable particles on neutrinos and BBN}
\label{sec:case-studies-qualitative}

Let us now qualitatively analyze the impact of the evolution of $Y$ particles on the dynamics of the MeV plasma. We will consider several aspects: properties of primordial neutrinos -- $N_{\rm eff}$, neutrino spectral distortions, the asymmetry in the energy distribution between neutrinos and antineutrinos, and the neutron-to-proton conversion, which sets the initial condition for BBN.

The purpose of this analysis is to understand the impact of metastable particles' dynamics on the population of neutrinos in simple terms. This way, it accompanies the numerical study of the evolution of the neutrino distribution function (sec.~\ref{sec:unintegrated}), used to obtain our main results -- figures~\ref{fig:cross-check-1},~\ref{fig:DeltaNeff-toy}, and~\ref{fig:Neff-scalar}.

\subsection{$N_{\rm eff}$ and Neutrino Spectral Distortions}
\label{sec:neff_distortions}

The effective number of relativistic neutrino species, $N_{\rm eff}$, is defined as
\begin{equation}
    N_{\text{eff}} = \frac{8}{7}\left(\frac{11}{4}\right)^{\frac{4}{3}} \frac{\rho_{\text{UR}} - \rho_{\gamma}}{\rho_{\gamma}} \Bigg|_{m_{\nu} \ll T \ll m_{e}},
\end{equation}
where $\rho_{\text{UR}}$ and $\rho_{\gamma}$ represent the energy densities of ultra-relativistic particles and photons, respectively. Under the assumption that neutrinos follow an equilibrium (Fermi-Dirac) distribution, $N_{\rm eff}$ effectively characterizes the neutrino population. However, deviations from thermal equilibrium can lead to a non-thermal neutrino distribution function, $f_{\nu}(p,t)$ and break this degeneracy.

In the $\Lambda$CDM framework, the value of $N_{\rm eff}$ is $N_{\rm eff}^{\Lambda\text{CDM}} \approx 3.04$~\cite{Mangano:2001iu,Bennett:2019ewm,Bennett:2020zkv,Akita:2020szl,Froustey:2020mcq,Cielo:2023bqp,Jackson:2023zkl,Drewes:2024wbw}, and the neutrino distribution closely resembles a Fermi-Dirac distribution with temperature $T_{\nu} \approx (4/11)^{1/3} T_{\gamma}$. Variations in $N_{\rm eff}$ and $f_{\nu}(p,t)$ influence the Universe's expansion rate and the neutron-to-proton conversion rates. Specifically, energetic neutrinos can efficiently convert protons to neutrons, thereby increasing the neutron-to-proton ratio beyond the $\Lambda$CDM prediction and enhancing primordial helium abundance. Additionally, distortions break the degeneracy between the neutrino energy and number densities, which may be important in the epoch when they become non-relativistic.

Without decays into metastable particles, there are two distinct scenarios:

\begin{enumerate}
    \item \textit{LLPs decaying solely into EM particles}: In this scenario, the evolution of the neutrino population may be approximately described in terms of the evolution of its temperature~\cite{EscuderoAbenza:2020cmq} (see a discussion in Sec.~\ref{sec:unintegrated}). The resulting deviation in the effective number of relativistic degrees of freedom is $\Delta N_{\rm eff} = N_{\rm eff} - N_{\rm eff}^{\Lambda\text{CDM}} < 0$, with the neutrino temperature $T_{\nu}$ being lower than in the standard case, $T_{\nu} < T_{\nu}^{\Lambda\text{CDM}}$ due to the heating of the EM plasma by the energy injection from the $X$ decays.
    
    \item \textit{LLPs decaying solely into neutrinos}: Decays into neutrinos with thermal energies $E_{\nu}\simeq 3T$ have the opposite effect compared to the pure EM decays: heating the neutrino plasma and increasing $N_{\rm eff}$. Decays into high-energy neutrinos ($E_{\nu} \gg 3T$) in MeV plasma have a qualitatively different impact. As detailed in~\cite{Boyarsky:2021yoh, Ovchynnikov:2024rfu, Ovchynnikov:2024xyd}, they reduce $N_{\rm eff}$, which arises from two main effects:
    \begin{itemize}
        \item Spectral distortions: high-energy neutrinos interact with thermal neutrinos, enhancing the high-energy tail and depleting the low-energy part of the spectrum. It is important since the rates of the neutrino-EM interaction grow with the energy of the particles.
        \item Instant thermalization of the EM plasma: any energy injection to the EM sector instantly thermalizes. Without distortions in the spectrum of $e^{\pm}$ particles, the net energy flow is shifted to the EM sector even when the ratio of the energy densities $\rho_{\nu}/\rho_{\text{EM}}$ reaches equilibrium value. As a result, this shift leads to $\Delta N_{\rm eff} < 0$, analogous to pure EM plasma heating.
    \end{itemize}
    As the LLP lifetime $\tau_X$ increases, high-energy neutrinos interact less with the EM plasma, reducing the energy transfer and mitigating the negative impact on $N_{\rm eff}$. For sufficiently large $\tau_X$, $\Delta N_{\rm eff}$ becomes positive.
\end{enumerate}
The scenario of LLPs decaying into metastable particles is even more nuanced. Given that $m_Y \gg 3T$, neutrinos from $Y$ decays are typically energetic and resemble the second scenario. However, at MeV temperatures, the decay probability $P_{\rm decay}^Y$ is a lot smaller than unity, meaning that $Y$ particles are more likely to annihilate or interact with nucleons before decaying, effectively suppressing neutrino injection. This behavior mimics pure EM heating. At lower temperatures, as $P_{\rm decay}^Y \to 1$, the situation transitions towards the mix between scenarios 1 and 2.

To qualitatively characterize the impact of the varying $P^{Y}_{\text{decay}}$ on $N_{\rm eff}$, we define the ratio
\begin{align}
    r_{\nu} &= \left. \frac{\rho_{\text{inj}, \nu}}{\rho_{\text{inj}}} \right|_{t=\infty},
    \label{eq:rnu} 
\end{align}
which represents the fraction of the LLP's total injected energy $\rho_{\text{inj}}$ allocated to neutrinos. $\rho_{\text{inj}}$ and the energy density injected into neutrinos $\rho_{\text{inj},\nu}$ evolve according to
\begin{equation}
    \frac{d\rho_{I}}{dt} + 4H\rho_{I} = \left( \frac{d\rho_{I}}{dt} \right)_{\text{source}},
\end{equation}
where
\begin{multline}
    \left( \frac{d\rho_{I}}{dt} \right)_{\text{source}} = \frac{m_{X}n_{X}}{\tau_{X}} \times \\
    \times \begin{cases}
        1, & \rho_{I} = \rho_{\text{inj}} \\
        \xi_{X\to \nu} + \sum_{y = Y,\bar{Y}} \frac{n_{y}}{n_{X}} P^{y}_{\text{decay}} \xi_{y\to \nu}, & \rho_{I} = \rho_{\text{inj},\nu}
    \end{cases}
    \label{eq:source}
\end{multline}
Here, $\xi_{A\to \nu}$ denotes the fraction of the $A$'s energy injected into the neutrino sector per decay:
\begin{equation}
    \xi_{A\to \nu} = \frac{1}{m_{X}} \sum_{j} \text{Br}_{A,j} \langle E^{(j)}_{\nu} \rangle,
    \label{eq:energy-to-nu}
\end{equation}
with $\text{Br}_{A,j}$ denoting the branching ratio of the $j$th decay mode of the particle $A$, and $\langle E^{(j)}_{\nu} \rangle$ mean energy of neutrinos injected in this decay. When calculating it, we assume that all metastable particles do not decay. As an example, for the decay channel $K^{+} \to \mu^{+} + \nu_{\mu}$, only the neutrino energy is accounted for, whereas the muon is dropped.

As is seen from the definition of $r_{\nu}$, we drop any interactions between neutrinos and electromagnetic particles. This is done for qualitative studies -- to concentrate on the impact of the dynamics of metastables.

The minimum value of $r_{\nu}$ occurs when $P_{\rm decay}^Y = 0$, implying that only direct $X$'s decays into neutrinos would contribute. Conversely, the maximum value is achieved when $P_{\rm decay}^Y = 1$, meaning that all mesons and muons decay:
\begin{equation}
    r_{\nu,0} = \frac{1}{m_{X}} \sum_{j} \text{Br}_{j} \cdot \overline{\langle E^{(j)}_{\nu} \rangle},
    \label{eq:rnu-0}
\end{equation}
where, unlike eq.~\eqref{eq:energy-to-nu}, we include the contribution from inevitable decays when calculating the mean neutrino energy, $\overline{\langle E^{(j)}_{\nu} \rangle}$. As a cross-check, the expression~\eqref{eq:source} (and hence $r_{\nu}$) should give exactly the same results as eq.~\eqref{eq:rnu-0} in the case $P_{\rm decay}^{y} = 1$. We confirm this in Figs.~\ref{fig:rnu-toy},~\ref{fig:rnu-scalar},~\ref{fig:rnu-HNL} in the limit of large $X$ lifetimes.

When neutrinos from the decay of $Y$ particles effectively decouple, the sign of $\Delta N_{\rm eff} = N_{\rm eff} - N_{\rm eff}^{\Lambda\text{CDM}}$ is determined by whether $r_{\nu,0}$ exceeds the ratio of neutrino to total energy densities in standard cosmology, which for temperatures $T\gtrsim m_{e}$ is
\begin{equation}
    q_\nu = \frac{\rho_{\nu}}{\rho_{\nu} + \rho_{\text{EM}}} = \frac{21}{43}.
    \label{eq:ratio-equilibrium}
\end{equation}
If $r_{\nu,0} > q_{\nu}$, then $\Delta N_{\rm eff}$ increases as $\tau_X$ becomes large ($\tau_{X} \gtrsim 1\,\text{s}$). Consequently, $\Delta N_{\rm eff}$ transitions from negative to positive values as $r_{\nu}$ approaches $r_{\nu,0}$.\footnote{Note that $q_{\nu}$ decreases after electron-positron annihilation, allowing for an additional sign change if the LLP decays at $T \lesssim m_e$.}

\subsection{Neutrino-antineutrino energy asymmetry}
\label{sec:asymmetry}

Generically, the evolution~\eqref{eq:system-Y},~\eqref{eq:Xn-dynamic-equilibrium} is not $Y-\bar{Y}$ symmetric due to the term describing the interactions with nucleons. The reason is that there are no anti-nucleons, which means that the generic interaction rate of $Y$ and $\bar{Y}$ does not have charge conjugation symmetry. This implies that, in general, $n_{Y}\neq n_{\bar{Y}}$, i.e., metastable particles and antiparticles evolve differently. This asymmetry translates to an asymmetry between neutrinos and antineutrinos via their decays. Let us discuss its qualitative aspects.

The asymmetry may be in number ($n_{\nu}\neq n_{\bar{\nu}}$) and energy distributions (meaning in particular that $\rho_{\nu}\neq \rho_{\bar{\nu}}$). In the first case, the net lepton charge $L_{\nu} \propto n_{\nu}-n_{\bar{\nu}}$ is generated in the neutrino sector, and the opposite charge $L_{l} = -L_{\nu}$ in the electron-positron sector. No sizeable $L_{\nu}$ is induced because we assume that the initial $X$ particle is electrically neutral and the baryon number is conserved. Indeed, the electric charge conservation means that independently of the microscopics of the $Y,\bar{Y}$ evolution, $L_{l}$ may occur only because of changing the yield of protons. The baryon number conservation implies that this change is bounded by $\eta_{B}\sim 10^{-9}$. Therefore, we may just assume that $n_{\nu} = n_{\bar{\nu}}$. 

However, the magnitude of the energy asymmetry is not bounded by this argument. First, even if conserving the number of neutrinos, decays of different $Y$s inject neutrinos with different energies. Namely, decays of kaons release neutrinos with energies as large as $E_{\nu,\text{max}} \approx m_{K}/2$, decays of pions result in the neutrinos with energy $E_{\nu,\text{max}}\approx 29\text{ MeV}$, whereas the maximal neutrino energy from muons decays is $E_{\nu,\text{max}}\approx m_{\mu}/2$. Second, some $Y$\!s, such as kaons, may interact with both protons and neutrons, as well as may or may not convert them, meaning that $\rho_{\nu}-\rho_{\bar{\nu}}$ may easily exceed the bound $n_{B}\times E_{\nu}$ coming from the number asymmetry. 

Let us now discuss the energy asymmetry in more detail. If only muons are injected, the nucleon interaction term may be neglected (see a discussion in Sec.~\ref{sec:muons}). If, in addition, the $X$ particle decays into the charged pions, it is important, and we need to analyze it further. Both $\pi^{+},\pi^{-}$ interact with nucleons; in addition, 
\begin{equation}\langle\sigma_{\mathcal{N}}^{\pi^{\pm}}v\rangle =\langle\sigma_{\mathcal{N}\to \mathcal{N}'}^{\pi^{\pm}}v\rangle,
\end{equation}
i.e., pions interact with nucleons solely via converting them (remind eq.~\eqref{eq:nucleon-interaction-pions}). Using this and utilizing the expression for the nucleon abundance $X_{n}$ from eq.~\eqref{eq:Xn-dynamic-equilibrium}, we find that the nucleon interaction terms for $\pi^{+},\pi^{-}$ are actually identical:
\begin{multline}
n_{\pi^{-}}\sum_{\mathcal{N}}n_{\mathcal{N}}\langle\sigma^{\pi}_{\mathcal{N}} v\rangle = n_{\pi^{+}}\sum_{\mathcal{N}}n_{\mathcal{N}}\langle\sigma^{Y}_{\mathcal{N}} v\rangle \\ = \frac{n_{\pi^{-}}n_{\pi^{+}}\langle \sigma^{\pi^{-}}_{p\to n}v\rangle\langle \sigma^{\pi^{+}}_{n\to p}v\rangle}{n_{\pi^{-}}\langle \sigma^{\pi^{-}}_{p\to n}v\rangle+n_{\pi^{+}}\langle \sigma^{\pi^{+}}_{n\to p}v\rangle}
\end{multline}
The situation is different when charged kaons are injected as well. There is an explicit asymmetry due to the interaction with nucleons: $K^{+}$ does not interact with nucleons in the MeV plasma, while $K^{-}$ participates in various processes with them: interacting with both $n$ and $p$, converting $n\leftrightarrow p$ as well as keeping the nucleon type the same (remind Sec.~\ref{sec:kaons}). As a result, more $K^{-}$ would disappear before decaying than $K^{+}$. Decays of $K^{+}$ would directly produce muon neutrinos and not antineutrinos. On the other hand, it means that we have more $\pi^{+},\mu^{+}$ particles, that decay into antineutrinos. 

Overall, this decay asymmetry may induce sizeable differences in the energy distributions of neutrinos and antineutrinos. The energy asymmetry may be split into the ranges $E_{\nu}>m_{\mu}/2$, to which only the $K$ decays contribute, and $E_{\nu}< m_{\mu}/2$, where the main sources are decays of muons and pions. The first domain is overabundant for neutrinos, whereas the second is for antineutrinos. We leave the quantitative study of this intriguing asymmetry development question for future work.

\subsection{Evolution of the $n/p$ ratio}

As was mentioned in Sec.~\ref{sec:metastable-dynamics}, injecting mesons into the primordial plasma significantly modifies the dynamics of the $n/p$ ratio $n_{n}/n_{p}$. Overall, the effect of the meson-driven $p\leftrightarrow n$ conversion is well-known~\cite{Reno:1987qw,Kohri:2001jx,Pospelov:2010cw,Boyarsky:2020dzc}, but let us describe it shortly. In $\Lambda$CDM, the $n/p$ ratio is suppressed by the Boltzmann exponent as far as weak interactions maintain chemical equilibrium between the neutrons and protons:
\begin{equation}
\frac{n_{n}}{n_{p}} \approx \exp\left[-\frac{m_{n}-m_{p}}{T}\right]
\label{eq:chemical-equilibrium}
\end{equation}
Once mesons are injected, they increase the ratio above the value~\eqref{eq:chemical-equilibrium}. This is mainly because meson-driven $p\leftrightarrow n$ conversion is thresholdless. The BBN constraint on LLP lifetimes may be imposed from the requirement on this enlarged ratio to relax to the $\Lambda$CDM value within the margin determined by the error in the primordial helium measurements~\cite{Boyarsky:2020dzc}. The meson-driven $p\leftrightarrow n$ conversion cross-section is orders of magnitude higher than the one for the weak conversion, and even exponentially suppressed amounts of mesons (at times $t\gg \tau_{X}$) completely drive the dynamics of the $n/p$ ratio. Because of this, the resulting constraint on the LLP's lifetime depends on the LLP's initial number density and the yield of mesons available for the conversion only logarithmically~\cite[eq.~(11)]{Boyarsky:2020dzc}.

Because of the same reason, the meson-driven $p\leftrightarrow n$ conversion typically dominates over other effects of LLPs on the dynamics of the $n/p$ ratio, including the modified expansion of the Universe and neutrino properties. For example, consider Heavy Neutral Leptons with lifetimes $\tau_{N}\simeq 0.02\,\text{s}$ and heavy enough to decay into mesons. While modifying $N_{\rm eff}$ at a percent level, they induce a huge change in the $n/p$ ratio due to mesons~\cite{Boyarsky:2020dzc,Sabti:2020yrt}.

The only modification of this picture due to our study comes from adding the meson annihilation processes. They suppress the yield of mesons available for the $p\leftrightarrow n$ conversion, eq.~\eqref{eq:Pnucl}. However, the suppression is maximum a factor of few (remind fig.~\ref{fig:probabilities-simple}), which would modify the BBN constraint in a minor way as it enters the logarithm.

\section{Case studies}
\label{sec:case-studies}

In this section, we consider the impact of the evolution of $Y$\!s on neutrinos for three models with LLPs $X$: A toy model adding a particle with constant abundance decaying into charged pions (Sec.~\ref{sec:toy}), Higgs-like scalars (Sec.~\ref{sec:scalars}), and Heavy Neutral Leptons (HNLs) (Sec.~\ref{sec:hnls}). We will discuss the mass and lifetime dependence utilizing two approaches: the qualitative one, by calculating the fraction of $X$'s energy directly injected into neutrinos, eq.~\eqref{eq:rnu}, and the quantitative one, for which we will accurately trace the evolution of neutrinos using the unintegrated Boltzmann approach. The former serves the purpose of demonstrating the impact of varying decay probability on neutrino injections (Figs.~\ref{fig:rnu-toy},~\ref{fig:rnu-scalar},~\ref{fig:rnu-HNL}), while the latter is used to obtain the main results of this paper -- Figs.~\ref{fig:DeltaNeff-toy} and~\ref{fig:Neff-scalar}.

\subsection{Toy model: LLPs decaying into pions}
\label{sec:toy}

Consider a toy model with the LLP $X$ decaying solely into charged pions. It means that in eq.~\eqref{eq:system-Y}
\begin{equation}
    N_{\pi^{\pm}}^{X} = 1, \quad N_{\mu,K}^{X} = 0
\end{equation}
As a result, no kaons are involved, but there still would be muons originating from the pion decay, remind eq.~\eqref{eq:decay-pion}. To make the analysis as transparent as possible, the $X$'s abundance is chosen to be a constant:
\begin{equation}
    \mathcal{Y}_{\text{LLP}} \equiv \left( \frac{n_{\text{LLP}}}{s}\right)_{T = 10\text{ MeV}} = 2\cdot 10^{-3}
    \label{eq:toy-abundance}
\end{equation}
It corresponds to the scenario where the particle $X$ was in thermal equilibrium and decoupled while still being relativistic in a broad range of masses and lifetimes. We consider the masses above the di-pion decay threshold $m_X > 2 m_\pi \approx 0.28\,\text{GeV}$. Regarding the lifetimes, following the discussion in Sec.~\ref{sec:approach}, we test the range $0.01\,\text{s} < \tau_X < 10\,\text{s}$.

Let us first discuss how $X$ would distribute its energy among the neutrino and the EM sectors. Upon decay, it will produce a pair of pions whose non-trivial evolution has been discussed before. Injection into the neutrino sector would occur only in the case of decay of $\pi^{\pm}$ producing a muon and a muon (anti)neutrino, eq.~\eqref{eq:decay-pion}; the resulting muon may then decay into neutrinos as well (eq.~\eqref{eq:decay-muon}).

If the pion and the subsequently produced muon would inevitably decay, roughly $r_{\nu,0}^{\pi} \approx 70\%$ of the pion mass would go to the neutrino sector. In this case, the injection into the EM sector is composed of the initial kinetic energy of the pion $(m_X-2\cdot m_\pi)/2$ and the EM part of the muon decay, which is approximately $30 \%$ of its mass. This gives us the maximal possible fraction of the energy of the $X$ particle directly injected in the neutrino sector:
\begin{equation}
    r_{\nu,0} \approx \frac{2m_{\pi}\cdot r_{\nu,0}^{\pi}}{m_{X}} 
\approx q_{\nu}\cdot \frac{388\text{ MeV}}{m_{X}},
    \label{eq:X_rnu_onlydecay}
\end{equation}
where $q_{\nu}$ is the $\Lambda$CDM ratio~\eqref{eq:ratio-equilibrium}. Provided that there are no interactions of neutrinos with the EM plasma, if $r_{\nu,0}$ exceeds this ratio (i.e., $m_{X}<388\text{ MeV}$), the correction $\Delta N_{\text{eff}}$ would be positive in the limit of large lifetimes (remind the discussion in the previous section).

\begin{figure}[h!]
    \centering 
    \includegraphics[width=0.45\textwidth]{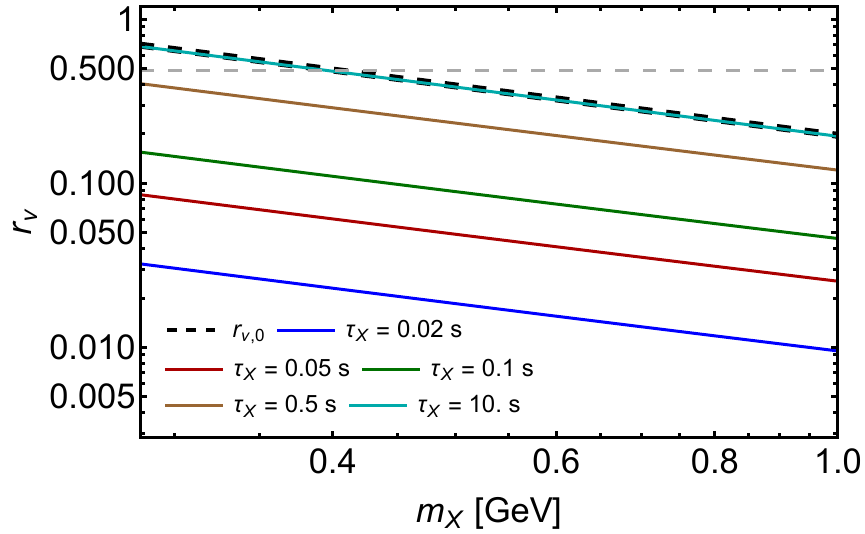} 
    \caption{The \textit{qualitative} impact of the dynamics of metastable particles on neutrinos for the model of a hypothetical particle $X$ decaying solely into a pair of the charged pions (see text for details). The plot shows the evolution of the fraction $r_{\nu}$ of the $X$'s energy directly injected into neutrinos, eq.~\eqref{eq:rnu}, as a function of the $X$'s mass $m_{X}$ and lifetime $\tau_{X}$ (shown by different solid lines). The gray dashed line shows the value $r_\nu = 21/43$, for which $\Delta N_{\text{eff}} = 0$ in the absence of the neutrino-EM interactions (see a discussion around eq.~\eqref{eq:ratio-equilibrium}). For the fixed  $X$ mass, $r_{\nu}(\tau_{X})\to 0$ at lifetimes $\tau_{X}\ll 1\,\text{s}$. Once lifetime increases, it gradually grows and, for $\tau_{X}\gg 1\text{ s}$, approaches the value $r_{\nu,0}$ (the dashed black line), which is when all pions and muons inevitably decay (eq.~\eqref{eq:X_rnu_onlydecay}). The pattern occurs since $\mu$s' and $\pi$s' scattering and annihilation processes, preventing them from releasing energy into neutrinos, become less efficient at lower temperatures (remind sec.~\ref{sec:simple}). The slope of the lines represents the increasing kinetic energy of the pion as a function of the LLP mass; it gets immediately transferred to the EM sector independently of the lifetime.}
    \label{fig:rnu-toy}
\end{figure}

The disappearance of the pions and muons because of annihilation and interaction with nucleons spoils this picture. Let us first estimate the impact of these effects qualitatively. Namely, we utilize the discussion of Sec.~\ref{sec:case-studies-qualitative} and calculate the quantity $r_{\nu}(m_{X},\tau_{X})$, see fig.~\ref{fig:rnu-toy}. At small lifetimes $\tau_X \lesssim 0.5$ s, $r_{\nu}\ll r_{\nu,0}$; this is because pions and muons produced in $X$ decays would prefer to disappear before decaying at $T\gtrsim 1\text{ MeV}$, remind sec.~\ref{sec:simple}. Therefore, decays into pions and muons affect the neutrino bath very similar to the solely electromagnetic decays. With the increase of the lifetime, more and more $Y$\!s would decay, and $r_{\nu}$ tends to the maximal possible value $r_{\nu,0}$.

\begin{figure}[h!]
    \centering
    \includegraphics[width=0.9\linewidth]{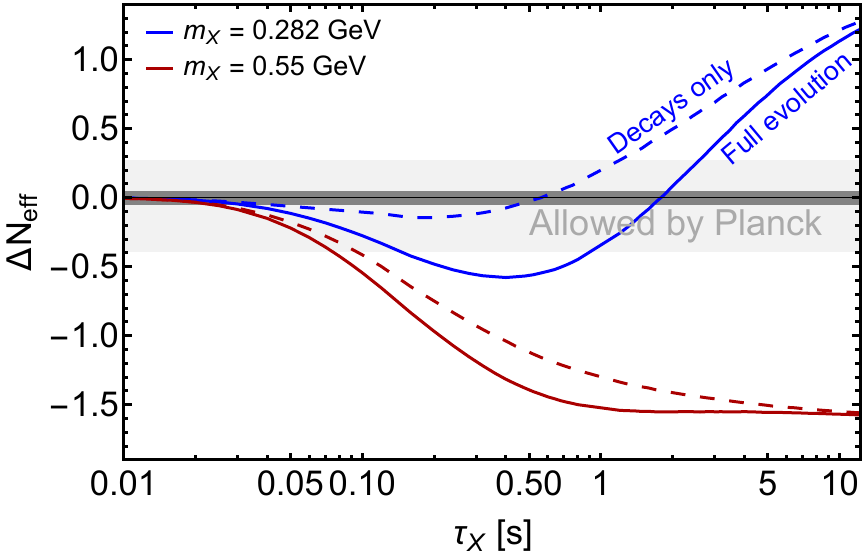} \\ \includegraphics[width=0.9\linewidth]{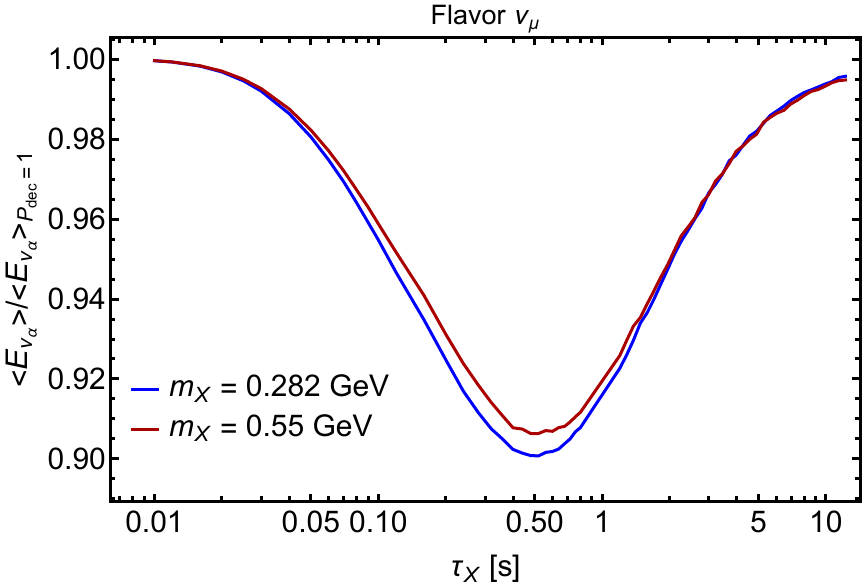} 
    \caption{The effect of the evolution of $Y$\!s on the properties of neutrinos for the same LLP model as in fig.~\ref{fig:rnu-toy}. The results are obtained using the unintegrated neutrino Boltzmann equation solver incorporating the dynamics of metastable particles, which we describe in Sec.~\ref{sec:unintegrated}. Two LLP masses are considered -- $m_{X} = 282\text{ MeV}$ and $550\text{ MeV}$. They represent the cases when the correction $\Delta N_{\text{eff}} = N_{\text{eff}} - N_{\text{eff}}^{\Lambda\text{CDM}}$ tends, correspondingly, to a positive and negative value in the limit of large lifetimes (see a discussion in Sec.~\ref{sec:toy}). To highlight the importance of the annihilation and interactions with nucleons, we consider two setups for each of the masses: the one that includes annihilation and interactions with nucleons (the realistic setup) and the one that includes solely decays and kinetic energy loss, which corresponds to the assumption $P_{\rm decay}^{Y} = 1$ used in all previous studies. \textit{Top panel}: the correction $\Delta N_{\text{eff}}$. The gray band represents the Planck 95\% CL constraints $N_{\text{eff}} = 2.99^{+0.33}_{-0.34}$~\cite{Planck:2018vyg}, whereas the black band shows the forecast of the accuracy of the measurements performed by the Simons Observatory, which we assume to be centered at $\Delta N_{\text{eff}} = 0$~\cite{SimonsObservatory:2018koc}. \textit{Bottom panel}: the ratio of the mean energies of the muon neutrinos in the realistic setup case to the setup $P_{\rm decay} = 1$, as a function of the $X$'s lifetime. The numerical noise in the domain of large lifetimes is caused by the precision limit of the Boltzmann solver.}
    \label{fig:DeltaNeff-toy}
\end{figure}

Now, let us calculate $N_{\text{eff}}$ as a function of the LLP's mass and lifetime for this toy model. For this purpose, we switch to the full unintegrated Boltzmann approach described in Sec.~\ref{sec:unintegrated}. The plot with $\Delta N_{\text{eff}}$ as a function of $X$ mass and lifetime is shown in the upper panel of fig.~\ref{fig:DeltaNeff-toy}. Two respresentative choices for the mass of $X$ are considered: $m_{X} = 282\text{ MeV}$, for which $r_{\nu,0}>q_{\nu}$ (and hence $\Delta N_{\text{eff}}$ would be positive at large lifetimes), and $m_{X} = 550\text{ MeV}$, for which $r_{\nu,0}<q_{\nu}$. To highlight the impact of the $Y$ evolution on the properties of neutrinos, we show the results for two setups -- the one assuming $P_{\text{decay}}^{Y} = 1$ (i.e., when the decays are inevitable), and the one including the full evolution of $Y$\!s, i.e. accounting for annihilations and interactions with nucleons (which we will call below the realistic setup).

The behavior of the curves in fig.~\ref{fig:DeltaNeff-toy} agrees with the qualitative discussion in Sec.~\ref{sec:case-studies-qualitative} and in this section. For the lifetimes $\tau_{X}\ll 10\,\text{s}$ and both masses, there are severe differences in $\Delta N_{\text{eff}}$ between the two setups. The realistic setup corresponds to a lower $\Delta N_{\text{eff}}$; this is expected since the decay of $Y$ particles injects more energy directly into the EM sector. In the limit of large lifetimes $\tau_{X}\to 10\,\text{s}$, the two results match, as annihilation and interactions with nucleons become irrelevant.

Note that our approach predicts a \textit{decrease} of $N_{\text{eff}}$ in the presence of the LLPs with the lifetimes $\tau_{X} \lesssim 1\text{ s}$, decaying into neutrinos of high energies $E_{\nu}\gg T$ at MeV temperatures. These results are aligned with the previous studies utilizing different unintegrated approaches~\cite{Ruchayskiy:2012si,Boyarsky:2021yoh,Rasmussen:2021kbf} (see also a discussion in refs.~\cite{Ovchynnikov:2024rfu,Ovchynnikov:2024xyd}).

To investigate the impact of the $Y$ disappearance further, let us consider the ratio of the mean neutrino energies after the electron-positron annihilation for these two setups; see the lower panel of fig.~\ref{fig:DeltaNeff-toy}. The setup with $P_{\text{decay}} = 1$ leads to higher neutrino energies, which is expected, as we have a more abundant high-energy neutrino tail.

\subsection{Higgs-like scalars}
\label{sec:scalars}

Let us now consider a particular model of long-lived particles. We start with Higgs-like scalars $S$~\cite{Boiarska:2019jym}. We will concentrate on the minimal model with the effective Lagrangian
\begin{equation}
    \mathcal{L} = \theta m_{h}^{2}hS + \mathcal{L}_{\text{kinetic}}
\end{equation}
Here, $h$ is the Higgs boson, and $\theta$ is the mixing angle, with $|\theta|\ll 1$. Due to the mass mixing, the scalars have a similar interaction pattern as $h$ (so Yukawa couplings to the SM fermions), with the couplings additionally suppressed by $\theta$. 

\begin{figure}[t!]
    \centering \includegraphics[width=0.45\textwidth]{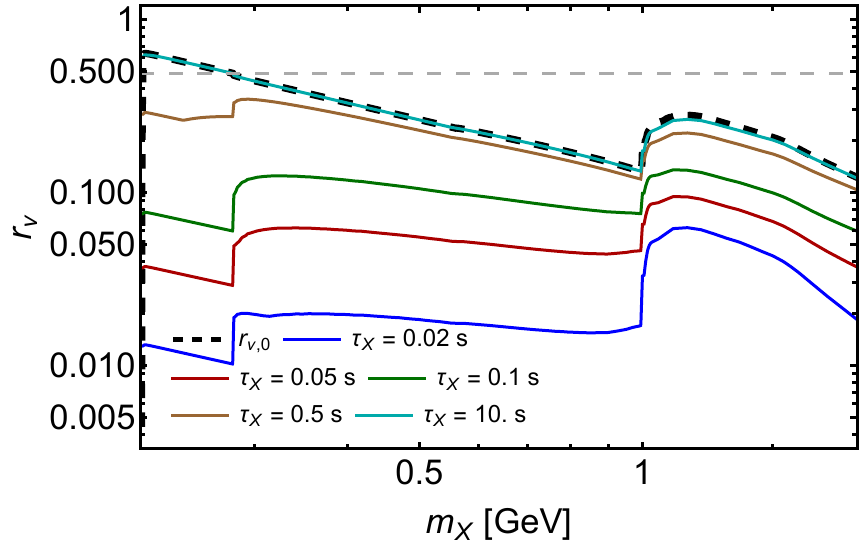} 
    \caption{The fraction of the energy directly injected into neutrinos in decays of Higgs-like scalar, see eq.~\eqref{eq:rnu} (remind also fig.~\ref{fig:rnu-toy}). The solid lines correspond to different scalar lifetimes, whereas the dashed black line is obtained under the usual assumption that all the metastable particles decay. The rapid change at the masses $m_{X} = 2m_{\mu},2m_{\pi},2m_{K}$ is caused by the opening of the decay into the pair of corresponding particles. The horizontal gray dashed line denotes the value of $r_{\nu,0}$ for which the injections would increase the neutrino-to-EM energy density ratio if assuming no interactions in the primordial plasma (see a discussion around eq.~\eqref{eq:ratio-equilibrium}).}
    \label{fig:rnu-scalar}
\end{figure}

The main decay modes of these scalars in the GeV mass range are two-body decays into particle-antiparticle pairs:
\begin{equation}
S\to e^{+}e^{-}/\mu^{+}\mu^{-}/\pi^{+}\pi^{-}/2\pi^{0}/K^{+}K^{-}/K_{L}K_{S},    
\end{equation}
with the decays into heavier particles dominating once they become kinematically possible. The fraction of energy injected into neutrinos $r_{\nu,0}$ by the scalar decays is shown in fig.~\ref{fig:rnu-scalar}. It is exactly zero for masses $m_{S}<2m_{\mu}$, because the only available scalar decay modes are into the EM particles. Then, it gets rapidly enhanced at $m_{S} = 2m_{\mu}$ and $m_{S} = 2m_{K}$ -- the mass thresholds where decays into two muons and kaons open up. In the domain of intermediate masses, it gradually decreases as the decay products have more and more kinetic energy that gets stored in the EM plasma.

The cosmological production and constraints of $S$ have been studied in~\cite{Fradette:2017sdd,Fradette:2018hhl}. Among the cosmological effects of the scalars, it studied the impact of the Higgs-like scalars on neutrinos. The analysis was simplified by considering a version of the integrated neutrino Boltzmann equation and assuming that $Y= \mu,\pi,K$ decay after thermalizing their kinetic energy. Under this approximation, $\Delta N_{\text{eff}}$ is determined by whether $r_{\nu,0}$ exceeds the quantity $q_{\nu}$ during decays of the scalar. This is the case in the region $2m_{\mu}< m_{S}\lesssim 2m_{\pi}$.

\begin{figure}[t!]
    \centering
    \includegraphics[width=\linewidth]{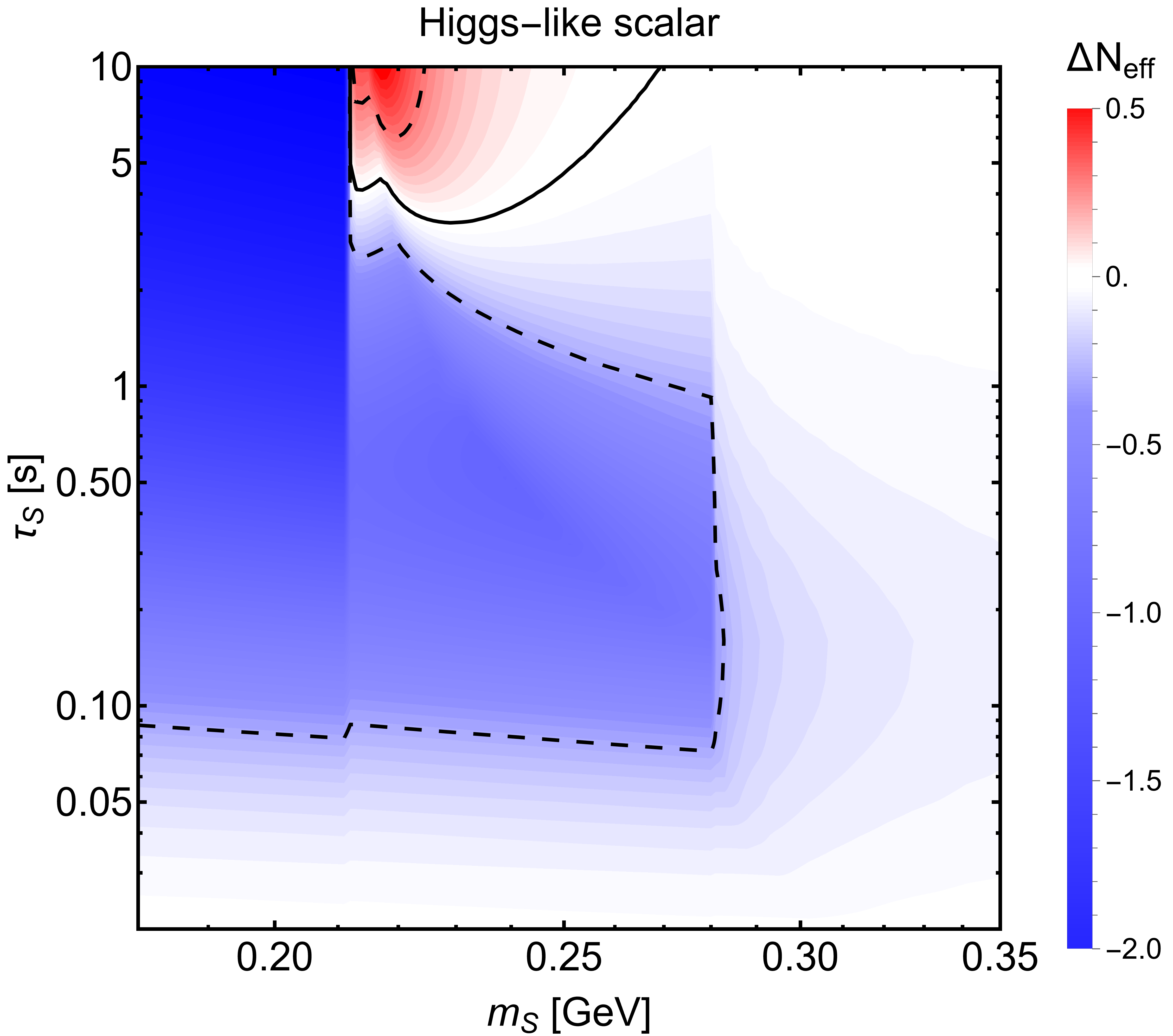}
    \caption{The effect of the presence of the Higgs-like scalars on the correction $\Delta N_{\text{eff}} = N_{\text{eff}}-N_{\text{eff}}^{\Lambda\text{CDM}}$, obtained by using the unintegrated Boltzmann equations from Sec.~\ref{sec:unintegrated}. The solid black line shows the parameter space where $\Delta N_{\text{eff}}$ = 0, whereas the dashed black lines denote the domain where $\Delta N_{\text{eff}}$ are beyond the lower and upper bounds of the $N_{\text{eff}}$ measurements as extracted from Planck measurements~\cite{Planck:2018nkj}. The change in the sign of $\Delta N_{\text{eff}}$ is driven by the dynamics of metastable particles produced by $S$'s decays. The decrease of the magnitude of $|\Delta N_{\text{eff}}|$ with the scalar mass is caused by the scaling of the scalar abundance $\mathcal{Y}_{S}(m_{S})\propto \Gamma^{-1}_{S}(m_{S},\theta = 1)$, where $\Gamma_{S}$ is the scalar decay width (see~\cite{Fradette:2017sdd} for details).}
    \label{fig:Neff-scalar}
\end{figure}

Let us now include the effects of mesons and muons evolution as well as momentum-dependent (unintegrated) Boltzmann equations for the neutrino distributions. In fig.~\ref{fig:rnu-scalar}, we show the mass-lifetime dependence of $r_{\nu}$ including the impact of annihilation and interactions. Similarly to the case of the toy model decaying into pions, the generic pattern is that $r_{\nu}(\tau_{S})\to 0$ for small scalar lifetimes and reaches $r_{\nu,0}$ for the lifetimes $\simeq 10\,\text{s}$. In particular, the ratio $r_{\nu}$ becomes less than $5\%$ (and so most of the scalar's energy goes to the EM sector) for the lifetimes $\tau_{S}\lesssim 0.05\,\text{s}$. $r_{\nu}$ jumps at $m_{S} = 2m_{\pi}$, which is caused by the opening of the di-pion decay channel. The pions (the main decay products in the mass range $2m_{\pi}<m_{S}<2m_{K}$) have a larger decay probability than the muons, which means that they have a higher chance to release energy into neutrinos than muons.
The behavior of $\Delta N_{\text{eff}}$ is shown in fig.~\ref{fig:Neff-scalar}. In the mass range $2m_{\mu}<m_{S}\lesssim 2m_{\pi}$, increasing the scalar lifetime, we see the transition between negative and positive changes in $N_{\text{eff}}$. It is caused by tending $r_{\nu}\to r_{\nu,0}>q_{\nu}$. At higher masses, $r_{\nu,0}<q_{\nu}$, so in any case, $\Delta N_{\text{eff}}$ remains negative. 

\subsection{Heavy Neutral Leptons}
\label{sec:hnls}

Let us now consider Heavy Neutral Leptons. The Lagrangian of HNLs has the form 
\begin{equation}
    \mathcal{L} = y_{\alpha} \bar{L}_{\alpha}\tilde{H}\text{HNL} + \text{h.c.},
\end{equation}
where $\alpha$ denotes the SM lepton generation, $L_{\alpha}$ is the corresponding left doublet, $y_{\alpha}$ is the Yukawa interaction coupling, while $\tilde{H} = i\sigma_{2}H^{*}$ is the Higgs doublet in the conjugated representation. 
Effectively, HNLs interact as heavy neutrinos, with the interaction coupling being suppressed by the mixing angle $U_{\alpha} \simeq y_{\alpha}v_{H}/m_{\text{HNL}}$, where $v_{H}$ is the Higgs VEV~\cite{Bondarenko:2018ptm}. We will consider the case of HNLs mixing with the muon neutrinos $\nu_{\mu}$, keeping in mind that the other cases are similar. 

Let us briefly discuss the production of HNLs. In high-temperature plasma, the mixing angle gets modified because of the thermal neutrino self-energy correction. In particular, in the plasma without the lepton asymmetry at temperatures $T\gtrsim 1\text{ GeV}$, the effective mixing angle is given by
\begin{equation}
    U^{2}_{\rm m}(T) \approx \frac{U^{2}}{\left[1+9.6\cdot 10^{-24}\left(\frac{T}{1\text{ MeV}}\right)^{6}\left(\frac{m_{\text{HNL}}}{150\text{ MeV}}\right)^{-2}\right]^{2}},
    \label{eq:hnl-mixing-angle-matter-effects}
\end{equation}
where $m_{\text{HNL}}$ is the HNL mass. The scaling of the HNL production rate with temperature is $\Gamma_{\text{int}} \sim G_{F}^{2}T^{5}U_{\rm m}^{2}$, with $G_{F}$ being the Fermi coupling. Comparing the HNL interaction rate with the Hubble rate $H$, we may establish whether HNLs entered the thermal equilibrium. Namely, the ratio $\Gamma_{\text{int}}/H$ is $\ll 1$ at high temperatures $T$ because of the suppression of $U_{m}(T)$, then reaches the peak value at $T_{\text{peak}} \approx 12\text{ GeV}\left( m_{\text{HNL}}/(1\text{ GeV})\right)^{1/3}\text{ GeV}$, and then starts decreasing, since $\Gamma_{\text{int}}$ drops with $T$ faster than $H$. If the rate-to-Hubble ratio at $T_{\text{peak}}$ is $<1$, HNLs never entered thermal equilibrium. 

\begin{figure}[t!]
    \centering
    \includegraphics[width=0.9\linewidth]{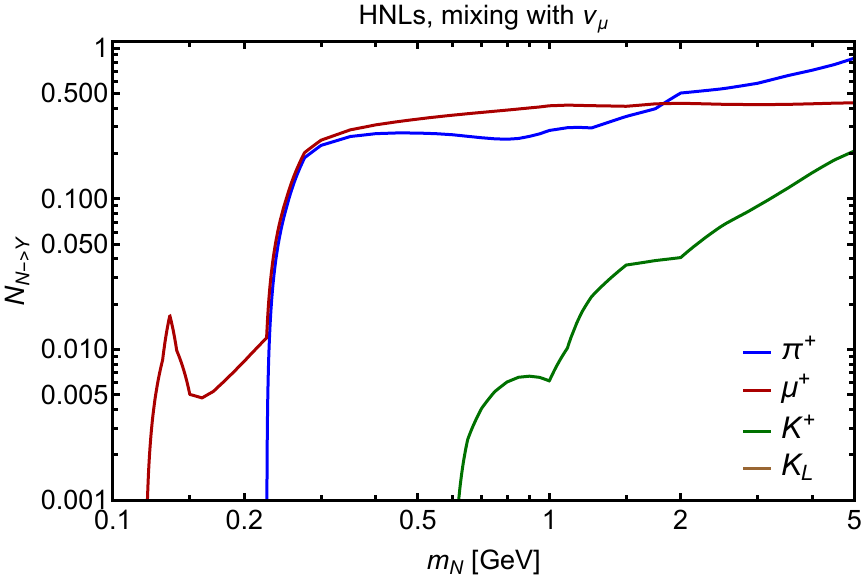}\\ \includegraphics[width=0.9\linewidth]{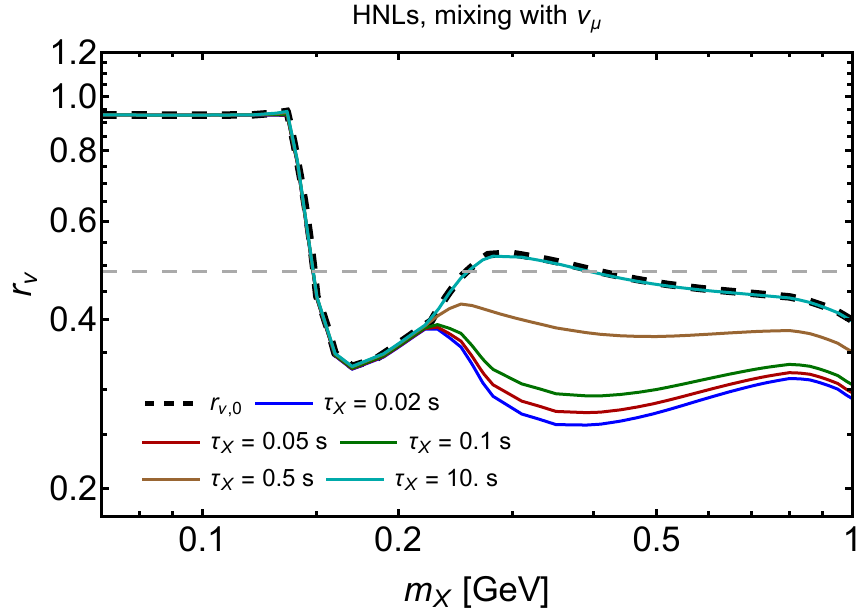} 
    \caption{The qualitative impact of the metastable particle dynamics on neutrinos for the case of HNLs coupled to the muon neutrinos. Top panel: the fractions of the metastable particles $Y = \mu^{\pm},\pi^{\pm},K^{\pm}$ per HNL decay. Bottom panel: the dependence of $r_{\nu}$ on the HNL lifetime. The minimal value of $r_{\nu}$ corresponds to the situation when all mesons and muons disappear without decaying; then, $r_{\nu}$ is saturated solely by direct decays into neutrinos.}
    \label{fig:rnu-HNL}
\end{figure}

We have calculated the HNL abundance following the approach similar to the one used in Ref.~\cite{Boyarsky:2020dzc}. To compute the kinematics of HNL decay products, we used the \texttt{SensCalc} package~\cite{Ovchynnikov:2023cry}, which we have modified to account for transferring of all the kinetic energy of the charged metastable particles to the EM plasma and forbid the mesons and muons to decay. We used the exclusive decays below the HNL mass $m_{\text{HNL}}\simeq 1\text{ GeV}$ and decays into jets above this mass, with showering and hadronization performed using \texttt{PYTHIA8}~\cite{Bierlich:2022pfr}. The amounts of the charged pions, muons, and kaons per HNL mass are shown in fig.~\ref{fig:rnu-HNL} (top panel).

Using the machinery described above, we computed the quantity $r_{\nu}$; see the same figure (bottom panel). Unlike the models considered above, the decay palette of the HNLs includes processes directly injecting neutrinos. They are 
\begin{equation}
\text{HNL} \to \nu_{\alpha}\bar{\nu}_{\alpha}\nu_{\beta}, \ \text{HNL} \to \text{hadrons}+\nu_{\beta},
\end{equation}
where ``hadrons'' denote either a single meson such as $\pi^{0}$ or a multi-meson state, depending on the HNL mass~\cite{Bondarenko:2018ptm}. Therefore, even if all the mesons and muons disappear without decaying, the quantity $r_{\nu}$ would be non-zero even for short HNL lifetimes $\tau_{\text{HNL}}\lesssim 0.1\,\text{s}$. However, the fraction to be injected by meson decays is still large, depending on the HNL mass.

Because of the presence of the direct decays into neutrinos, computing the impact of HNLs on the primordial neutrinos is much more complicated: the traditional approaches of solving the neutrino Boltzmann equation based on the discretization of the comoving momentum space would take a too large amount of time to evolve the neutrino distribution function. We will return to this in future work. Nevertheless, fig.~\ref{fig:rnu-HNL} shows the importance of careful tracing of the evolution of the metastable particles in the HNL case. 

\section{Conclusions}
\label{sec:conclusions}

Many new physics scenarios introduce long-lived heavy particles $X$, that decay into \textit{metastable} Standard Model (SM) particles $Y$, such as muons, charged pions, and kaons. Examples of the $X$ particles include Higgs-like scalars, dark photons, axion-like particles, and others. The lifetimes of the $Y$ particles are sufficiently long to allow numerous interactions with components of the primordial plasma, including electromagnetic particles and nucleons. These interactions significantly modify the evolution of $Y$ abundances and, consequently, affect the properties of primordial neutrinos.

In this work, we conducted a detailed study of $Y$ particle evolution, incorporating processes such as annihilation with antiparticles, interactions with nucleons, elastic electromagnetic scattering, and decays (see Sec.~\ref{sec:Y}). Notably, annihilation processes are examined here for the first time, while interactions with nucleons have previously been considered only regarding Big Bang Nucleosynthesis (BBN) and not when studying the impact on neutrinos.

We have outlined a two-step scheme to trace the dynamics of $Y$ particles and their impact on neutrinos in Sec.~\ref{sec:approach}.
The first step is to analyze the coupled dynamics of $Y$ particles and nucleon densities. We have developed a systematic approach based on the system of the integrated Boltzmann equations on their number densities (Sec.~\ref{sec:metastable-dynamics}) and incorporated the method in a public \texttt{Mathematica} code. We have demonstrated that at MeV temperatures, $Y$ particles predominantly annihilate or interact with nucleons rather than decay (Sec.~\ref{sec:simple}).

As step two, we have incorporated the $Y$ dynamics in an unintegrated Boltzmann solver for the neutrino momentum distribution functions, see Sec.~\ref{sec:unintegrated}. It may be used to study a broad range of LLPs, including those decaying into metastable particles, as well as neutrinos or purely electromagnetic particles. Being public and simple in use, it allows the scientific community to robustly trace the evolution of neutrinos in the presence of new physics, which is especially important in light of future CMB observations. 

The non-trivial dynamics of $Y$ particles substantially alters the influence of new physics on neutrino properties, which we discuss in qualitative terms in Sec.~\ref{sec:case-studies-qualitative}. Specifically, if $Y$ particles decay, a significant fraction of their mass energy is transferred to the neutrino sector, inducing spectral distortions. Conversely, if they disappear without decaying, their energy is instead fully injected into the electromagnetic sector. Additionally, the differential decay rates of kaons and antikaons lead to asymmetries in the energy distributions of neutrinos and antineutrinos, which may persist if injections occur during neutrino decoupling (Sec.~\ref{sec:asymmetry}). A comprehensive analysis of this intriguing question is left for future work.

We applied the combined methodology of Secs.~\ref{sec:unintegrated} and~\ref{sec:case-studies-qualitative} to specific models with LLPs decaying into metastable particles: a toy model with pions (Sec.~\ref{sec:toy}), Higgs-like scalars (Sec.~\ref{sec:scalars}), and Heavy Neutral Leptons (Sec.~\ref{sec:hnls}). Our findings reveal significant deviations from previous studies assuming the inevitability of $Y$ decays, including changes in both the magnitude and sign of $\Delta N_{\text{eff}}$ and alterations in the neutrino distribution functions, as illustrated in Figs.~\ref{fig:DeltaNeff-toy} and~\ref{fig:Neff-scalar}. 

In summary, our results provide a deeper understanding of how long-lived particles influence the neutrino population in the Early Universe.

\bigskip

\textbf{Acknowledgements.} This work has received support by the European Union’s Framework Programme for Research and Innovation Horizon 2020 under grant H2020-MSCA-ITN-2019/860881-HIDDeN, JSPS Grant-in-Aid for Scientific Research KAKENHI Grant No.~24KJ0060. K.A is grateful to Maksym Ovchynnikov and Thomas Schwetz for the hospitality during the stay at the Institute for Astroparticle Physics, KIT. The authors thank Miguel Escudero for carefully reading the manuscript and for providing useful comments.

\newpage

\onecolumngrid 

\appendix

\section{\texttt{Mathematica} code for the evolution of metastable particles}
\label{app:code}

In this Appendix, we discuss the \texttt{Mathematica} code that traces the evolution of the metastable particles in the presence of the decaying LLPs $X$; it is available on \faGithub \cite{GitHub-metastable} and Zenodo~\cite{Zenodo}. The \texttt{Zenodo} repository also contains pre-computed data for some LLP models. 

The central notebook is \texttt{main.nb}. Once launching its initialization cells, it first calls secondary notebooks with all the necessary definitions. The secondary notebooks are located in the folder \textit{codes/Secondary particles evolution}. They are: \texttt{parameters-functions.nb}, defining various parameters, analytic functions, and the list of the metastables \texttt{Yparticleslist}; \texttt{cross-sections.nb}, containing the calculations of various interaction rates involving $Y$\!s; \texttt{universe-simplified-dynamics.nb}, containing a simplified description of the thermodynamics following the approach of~\cite{EscuderoAbenza:2020cmq}; \texttt{evolution-Ys.nb}, which brings all the processes with $Y$\!s altogether and defines the system of the Boltzmann equations on the $Y$\!s' number densities, depending of various options and properties of the $X$ particles; and \texttt{final-system.nb}, which uses these codes to calculate the impact of the decaying LLPs with $Y$ decay products. Apart from that, the folder \textit{SM\_Rates} contains useful definitions such as effective Lagrangians, tabulated energy densities of electrons, and oscillation probabilities.

Once all secondary codes are called, users may use the main notebook to study various physics cases. As an input for the model, the code requires various properties. The input for the implemented models is stored in the section \textit{LLP input}. Each of its sub-sections is dedicated to a separate model. 

For the given model \texttt{LLP} and \texttt{mass}, \texttt{$\tau$} staying for the LLP's mass and lifetime, the main definitions are: 
\begin{itemize}
    \item \texttt{$\tau$LLP[LLP,mass,coupling]}, which describes the dependence of the lifetime on LLP's mass and coupling;
    \item \texttt{nLLPini[LLP, mass, $\tau$]}, which is the number density of the LLP in the units of $\text{GeV}^{3}$ at $T = 20\text{ MeV}$, $n_{\text{X,ini}}$; the code assumes that the LLPs are already decoupled at that epoch. 
    \item \texttt{\{$\xi$to$\nu$[LLP,$\nu_{e}$,mass],$\xi$to$\nu$[LLP,$\nu_{\mu}$,mass],$\xi$to$\nu$[LLP,$\nu_{\tau}$,mass]\}}, which are the mass-dependent fractions of the LLP's mass injected \textit{directly} in the neutrino sector, the flavor $\nu_{\alpha}$.
    \item \texttt{NtoY[LLP,Y, mass]} -- the amount of the $Y$ particles produced per LLP's decay. It is defined by eq.~\eqref{eq:NY}.
    \item \texttt{BrTo2$\nu$[LLP, "$\nu_{e}$", mass], BrTo2$\nu$[LLP, "$\nu_{\mu}$", mass], BrTo2$\nu$[LLP, "$\nu_{\tau}$", mass]}, which is the branching ratio of decays $X\to \nu\bar{\nu}$.
    \item \texttt{EnergyFractionsTo$\nu$[LLP, "Total", mass]}, which is the \emph{total} fraction injected into neutrinos, assuming that all $Y$ particles inevitably decay.
\end{itemize}

Note that the code is more generic than the unintegrated Boltzmann code. For example, it may consider any LLP with any decay mode into neutrinos, $K_{L}$\!s, etc.

The section \textit{Generating the evolution of Ys} is devoted to generating the data for the grid of masses and lifetimes of the given LLPs: \texttt{MassGrid[LLP]}, \texttt{lifetimeGrid[LLP]}, defined in section \textit{Mass and lifetime grids} for each LLP. Subsection \textit{Launching for mass and lifetime grids} launches the system of equations for the given LLP model, mass and lifetime grids. This is done with the help of the routine \texttt{exportBlockFullData[IfWritingOutputForBoltzmann]}, where the parameter \texttt{IfWritingOutputForBoltzmann} may be \texttt{True} (if one wants to prepare output for the unintegrated Boltzmann solver) or \texttt{False} (if does not). For each mass and lifetime, this routine launches 

\texttt{mergedFunction[LLP, mass, $\tau$, DecayOnly]} 

where the parameter \texttt{DecayOnly} (\texttt{True} or \texttt{False}) controls whether including annihilation and interactions with nucleons, and returns the following data:

\texttt{\{LLP parameters, tabulated decay probabilities of $Y$s\}}

where \texttt{LLP parameters} is the row

\texttt{\{mass,$\tau$,$n_{\text{X,ini}}$,$N_{\text{eff}}$,$r_{1}$,$r_{\nu}$,$r_{3}$,$N_{\mu^{+}}^{X}$,$N_{\pi^{+}}^{X}$,$N_{K^{+}}^{X}$,$N_{K_{L}}^{X}$,$r_{\nu,0}$,$\text{Br}_{X\to \nu_{e}\bar{\nu}_{e}}$,$\text{Br}_{X\to \nu_{\mu}\bar{\nu}_{\mu}}$,$\text{Br}_{X\to \nu_{\tau}\bar{\nu}_{\tau}}$\}}

Here, the value of $N_{\text{eff}}$ is obtained via the integrated approach of Ref.~\cite{EscuderoAbenza:2020cmq}, $r_{1}$ is the cumulative fraction of the total energy density injected by LLP to the total energy density of the Universe, $r_{3}$ is the ratio of the energy density injected into the neutrino sector to the total neutrino energy density. The quantities $N_{Y}^{X}$ are defined around eq.~\eqref{eq:system-Y}, while $r_{\nu},r_{\nu,0}$ are given by eqs.~\eqref{eq:rnu},~\eqref{eq:rnu-0}. The tabulated decay probabilities are provided in the form \texttt{\{time in seconds, $P_{\text{decay}}^{Y}$(time)\}}.

The routine \texttt{exportOutputForCluster[LLP]} prepares the data for the unintegrated Boltzmann solver for the neutrino Boltzmann equation from sec.~\ref{sec:unintegrated}.

There are also notebooks \texttt{plot-numbers.nb} and \texttt{plot-distributions.nb}. The former imports the datasets prepared by the integrated and unintegrated approaches and makes plots (subsection \textit{Plots}). The imported data has the names \texttt{OutputLLPintegrated[LLP]} and \texttt{OutputLLPUnintegrated[LLPsel]}, correspondingly for the output of the notebook and the Boltzmann solver. The latter is pre-computed for some models and may be found in the associated \texttt{Zenodo} repository. Each subsection (e.g., \textit{$r_{\nu}$}), contains definitions needed to make a plot for each model (say, \texttt{\{mminPlot[LLP, "$r_{\nu}$"], mmaxPlot[LLP, "$r_{\nu}$"]\}} defines the LLP mass range for the $r_{\nu}$ plot), as well as the code making the plot itself. 

The notebook \texttt{plots-distributions.nb} makes plots with the neutrino momentum distributions and average energies calculated using the unintegrated code.

\section{Approach of integrated Boltzmann equations}
\label{app:integrated}

According to the integrated approach to solve the neutrino Boltzmann equations~\cite{Escudero:2018mvt,EscuderoAbenza:2020cmq}, we need to solve the following system in order to trace the neutrino distribution:
\begin{equation}
\begin{cases}
    \frac{d\rho_{\nu_{\alpha}}}{dt}+4H\rho_{\nu_{\alpha}} =\left(\frac{d\rho_{\nu_{\alpha}}}{dt}\right)_{X}+\left(\frac{d\rho_{\nu_{\alpha}}}{dt}\right)_{\text{EM}\leftrightarrow \nu_{\alpha}}, \\ \frac{d\rho_{\text{EM}}}{dt}+
3H(p_{\text{EM}}+\rho_{\text{EM}})= \left(\frac{d\rho_{\text{EM}}}{dt}\right)_{X}-\sum_{\alpha}\left(\frac{d\rho_{\nu_{\alpha}}}{dt}\right)_{\text{EM}\leftrightarrow \nu_{\alpha}}, \\ H = \frac{\dot{a}}{a} = \sqrt{\frac{8\pi G}{3}(\rho_{X}+\rho_{\gamma}+\rho_{e}+\sum_{\alpha}\rho_{\nu_{\alpha}})}, \\ \rho_{X} \approx \left(\frac{a(t_{0})}{a(t)}\right)^{3}n_{X,0}m_{X}\exp\left[-\frac{t-t_{0}}{\tau_{X}}\right],
\end{cases}
    \label{eq:system-integrated}
\end{equation}
Here, $\rho_{\text{EM}} = \rho_{\gamma}+\rho_{e}$ is the energy density of the EM particles, with the electron-positron and photon components
\begin{equation}
    \rho_{\gamma} = \frac{1}{\pi^{2}}\int dE\frac{E^{3}}{\exp\left(\frac{E}{T}\right)-1}, \quad \rho_{e} = \frac{2}{\pi^{2}}\int dE \frac{E^{2}\sqrt{E^{2}-m_{e}^{2}}}{\exp\left(\frac{E}{T}\right)+1}
\end{equation}
$T$ is the temperature of the EM plasma. The pressure of the EM plasma is $p_{\text{EM}} = p_{e}+p_{\gamma}$, where
\begin{equation}
    p_{e} = \frac{2}{3\pi^{2}}\int dE \frac{(E^{2}-m_{e}^{2})^{\frac{3}{2}}}{\exp\left(\frac{E}{T}\right)+1}, 
    \quad p_{\gamma} = \frac{1}{3\pi^{2}}\int dE \frac{E^{3}}{\exp\left(\frac{E}{T}\right)-1} = \frac{\rho_{\gamma}}{3}
\end{equation}
$\rho_{\nu_{\alpha}}$ is similar for $\rho_{e}$, but with the extra prefactor $1/2$ and in the limit of zero mass, so that there is a simple analytic relation between the energy density and temperature $\rho_{\nu_{\alpha}} = 
\frac{7\pi^{2}T_{\nu_{\alpha}}^{4}}{120}$. Next, $\left(d\rho_{\nu_{\alpha}}/dt\right)_{X}$, $\left(d\rho_{\text{EM}}/dt\right)_{X}$ are the source terms due to injections of the particles by $X$'s decays.

$(d\rho_{\nu_{\alpha}}/dt)_{\text{EM}\leftrightarrow \nu_{\alpha}}$ is the energy exchange rate between neutrinos and the EM plasma~\cite{EscuderoAbenza:2020cmq}:
\begin{equation}
    \left(\frac{d\rho_{\nu_{\alpha}}}{dt}\right)_{\text{EM}\leftrightarrow \nu_{\alpha}} = \sum_{\beta}\langle P_{\beta\alpha}\rangle(T_{\text{EM}},\langle E_{\nu}\rangle)\frac{\delta \rho_{\nu_\beta}}{\delta t},
\end{equation}
with $\langle P_{\beta\alpha}\rangle$ being the neutrino oscillation probabilities from eq.~\eqref{eq:BE} but evaluated for the mean neutrino energy $\langle E_{\nu}\rangle = 3.15T_{\nu}$.

$\frac{\delta \rho_{\nu_\beta}}{\delta t}$ is the evolution rate of the individual neutrino flavor; we discuss it in the subsection below.

Finally, $t_{0}$ is some initial moment of time at which we define the $X$'s initial number density $n_{X,0} \equiv n_{X}(t_{0})$. It is taken as $T(t_{0}) = 20\text{ MeV}$.

Now, let us specify the source terms and energy transfer rates. Having the decay probabilities of the mesons and muons~\eqref{eq:Pdec}-\eqref{eq:Pnucl}, we may define the following source term for neutrinos:
\begin{equation}
\left(\frac{d\rho_{\nu_{\alpha}}}{dt}\right)_{\text{source}} = \frac{n_{X}}{\tau_{X}}\sum_{\beta}\langle P_{\beta\alpha}\rangle(T_{\text{EM}},\langle E_{\nu_{\beta}}\rangle)\bigg(\langle E_{X\to \nu_{\beta}}\rangle +\sum_{y = Y,\bar{Y}}\frac{n_{y}}{\tau_{y}}P^{y}_{\text{decay}}\langle E_{y\to \nu_{\beta}}\rangle\bigg),
    \label{eq:neutrinoSources}
\end{equation}
where 
\begin{equation}
    \langle E_{A\to \nu_{\alpha}}\rangle = \sum_{i}\text{Br}(A\to \nu_{\alpha})\cdot \langle E^{(i)}_{A\to \nu_{\alpha}}\rangle
   \label{eq:energy-to-nu-1}
\end{equation}
is the mean energy injected in the neutrino sector by decays of a particle $A$, while $\langle E_{y\to \nu_{\alpha}}\rangle$ is the mean energy per $y$ decay assuming that all $y'$ appearing per its decay are stable. To calculate it, we take into account only primary decays; i.e., for $K^{+}\to \mu^{+}+\nu_{\mu}$ we do not subsequently decay $\mu^{+}$, as it may disappear. We evaluate these energy fractions using the decay module of the code \texttt{SensCalc}~\cite{Ovchynnikov:2023cry}. The EM source term has the form
\begin{equation}
\left(\frac{d\rho_{\text{EM}}}{dt}\right)_{\text{source}} = \frac{n_{X}}{\tau_{X}}(m_{X}-\langle E_{X\to \nu}\rangle),
\end{equation}
assuming that the $X$ particle decays at rest.

The neutrino energy transfer rates are
\begin{align}
\frac{\delta \rho_{\nu_e}}{\delta t} &\approx \frac{G_F^2}{\pi^5} \left[ 4 \left(g_{eL}^2 + g_{eR}^2\right) F(T_\gamma, T_{\nu_e}) + F(T_{\nu_\mu}, T_{\nu_e}) + F(T_{\nu_\tau}, T_{\nu_e}) \right],\\
\frac{\delta \rho_{\nu_\mu}}{\delta t} &\approx \frac{G_F^2}{\pi^5} \left[ 4 \left(g_{\mu L}^2 + g_{\mu R}^2\right) F(T_\gamma, T_{\nu_\mu}) - F(T_{\nu_\mu}, T_{\nu_e})+F(T_{\nu_\tau}, T_{\nu_\mu}) \right], \\
\frac{\delta \rho_{\nu_\tau}}{\delta t} &\approx \frac{G_F^2}{\pi^5} \left[ 4 \left(g_{\mu L}^2 + g_{\mu R}^2\right) F(T_\gamma, T_{\nu_\tau}) - F(T_{\nu_\tau}, T_{\nu_e})-F(T_{\nu_\tau}, T_{\nu_\mu}) \right], 
\end{align} 
with $G_F\approx 1.167\cdot 10^{-5}\text{ GeV}^{-2}$ being the Fermi's constant,
\begin{equation}
g_{eL} = 0.727, \quad g_{eR} = 0.233, \quad g_{\mu L} = -0.273, \quad g_{\mu R} = 0.233,
\end{equation}
and the function
\begin{equation}
F(T_{x},T_{\nu_{\alpha}}) = 32 f_a^{\text{FD}} \mathcal{G}^{\nu_{\alpha}}_{x,\text{ann}}(T_{x})\left( T_{x}^9 - T_2^9 \right) + 56 f_s^{\text{FD}} \mathcal{G}^{\nu_{\alpha}}_{x,\text{scatt}}(T_{x}) T_{x}^4 T_{\nu_{\alpha}}^4 \left( T_{x} - T_{\nu_{\alpha}} \right)
\end{equation}
describes the temperature-dependent part of the energy exchange itself. Its analytic part is computed using the Maxwell-Boltzmann approximation and neglects the electron mass. To account for the Fermi-Dirac statistics, the following factors are introduced: $f_a^{\text{FD}} = 0.884$, $f_s^{\text{FD}} = 0.829$. To account for the electron mass, we take the interpolations of the corrections $\mathcal{G}^{\nu_{\alpha}}_{x,\text{ann}}(T_{x})$, $\mathcal{G}^{\nu_{\alpha}}_{x,\text{scatt}}(T_{x})$ from the repository accompanying~\cite{EscuderoAbenza:2020cmq}; they are non-unit only if $x$ represents the EM particle.

\section{Details of unintegrated Boltzmann equation implementation}
\label{app:unintegrated}

We incorporate the system~\eqref{eq:BE},~\eqref{eq:CE} in a \texttt{python} code based on ref.~\cite{Akita:2020szl}.\footnote{Available on \faGithub \cite{GitHub-solver}.} As an input, it requires the LLP mass and lifetime, the number density at some initial temperature, and the tabulated decay probabilities. Then, it evolves the neutrino population for a broad range of LLPs $X$ decaying into muons, pions, and neutrinos via a 2-body decay $X\to \nu\bar{\nu}$. These include Higgs-like scalars~\cite{Boiarska:2019jym}, generic pseudoscalars such as axion-like particles with various coupling schemes~\cite{Beacham:2019nyx,Bauer:2020jbp,Bauer:2021mvw,Aloni:2018vki,DallaValleGarcia:2023xhh}, particles coupled to quark currents like dark photons and $B-L$ mediators~\cite{Ilten:2018crw}, and majorons. Other models, such as HNLs and neutralinos, will be added in the future.

We numerically solve eqs.~\eqref{eq:BE} and~\eqref{eq:CE} with the following dimensionless variables: 
\begin{equation}
x=m_ea, \quad y=pa, \quad z =aT,
\end{equation}
approximately normalizing $z\rightarrow 1\ (a\rightarrow 1/T)$ at high-temperature limit (slight deviations from $z_{\text{in}} = 1$ are due to the entropy conservation
of electromagnetic plasma, neutrinos, and anti-neutrinos~\cite{Dolgov:1998sf}). $x,y,z$ characterize time, momentum, and the EM temperature, respectively. 

To solve the system of ODEs~\eqref{eq:BE}, we use the \texttt{RK45} method in \texttt{solve\_ivp} distributed in \texttt{scipy}~\cite{2020SciPy-NMeth}. We linearly discretize the neutrino momentum grid $y_i$ using 301 grid points with $y_{\rm min}=0.01$ and $y_{\rm max}={\rm max}[a_{\rm stop}E_{\nu,\,\text{max}}, 40\, {\rm MeV}]$. Here, $E_{\nu,\,\text{max}}$ is the maximal energy of injected neutrinos (depending on the LLP, it may, for example, be half the LLP mass or half the muon mass); $a_{\rm stop}$ is the scale factor after which further injections are neglected. It is chosen to correspond to a simulated time of $10\tau_X$, when it is reasonable to assume that essentially all LLPs will have decayed. Since $a_{\rm stop}$ needs to be known a priori, it is calculated using a tabulated relation $a(t)$ based on the case when LLPs are absent.
Finally, the integration is performed via a summed Simpson's rule. When we compute the integrations inside eq.~\eqref{eq:CE}, we linearly discretize the EM grid $y^{\rm QED}\in [0.01\ 20]$ using 81 grid points. It is enough, as the EM plasma is rapidly thermalized.

The dynamics of the metastable particles, muons and pions, is incorporated by relating their instant number densities to the number density of LLP and utilizing the decay probabilities~\eqref{eq:Pdec} as obtained in this study. Namely, in the source term, apart from the direct LLP decays into neutrinos (see, e.g.,~\cite{Sabti:2020yrt}), we include (see Appendix~\ref{app:source_BE} for the derivation)
\begin{multline}
   \left(\frac{df_{\nu_{\alpha}}(p)}{dt}\right)_{\text{source}} = \frac{2\pi^2}{p^2}\biggl|\frac{dn_X}{dt}\biggl|\times \bigg\{ N_{\pi}^XP_{\text{decay}}^{\pi}\bigg[\langle P_{\alpha \mu}\rangle\cdot \delta\left(E_{\nu} -\frac{m_{\pi}^{2}-m_{\mu}^{2}}{2m_{\pi}} \right) + P_{\text{decay}}^{\mu}(\langle P_{\alpha \mu}\rangle \mathcal{F}_{\nu_{\mu}}(p)+\langle P_{\alpha e}\rangle \mathcal{F}_{\nu_{e}}(p))\bigg] \\ + N_{\mu}^XP_{\text{decay}}^{\mu}[\langle P_{\alpha \mu}\rangle \mathcal{F}_{\nu_{\mu}}(p)+\langle P_{\alpha e}\rangle \mathcal{F}_{\nu_{e}}(p)]\bigg\},
   \label{eq:source-discretization}
\end{multline}
Here 
\begin{align}
    \biggl|\frac{dn_X}{dt}\biggl|=\frac{n_X(t)}{\tau_X}
\end{align}
is the differential rate of LLP's decay and $n_X(t)$ is given by eq.~\eqref{eq:nX}. $\delta(\dots)$ is the Dirac $\delta$ function. The first squared brackets describe the contribution from LLPs decaying into pions. It includes the direct decay $\pi^{+}\to \mu^{+}\nu_{\mu}$ into muon neutrinos (resulting the energy distribution to a $\delta$ function) and the secondary muon decay, which is $\mu^{+}\to e^{+}\nu_{e}\bar{\nu}_{\mu}$; the functions $\mathcal{F}_{\mu_{\alpha}}(p)$ are energy distributions of the neutrinos. The second brackets are from LLPs decaying into muons. 
The decay probabilities entering eq.~\eqref{eq:source-discretization} are taken as input from the tabulated files generated by our \texttt{Mathematica} code. Further details of usage may be found on the GitHub page.

Let us briefly discuss the limitations of the method. It utilizes the discretization approach from~\cite{Hannestad:1995rs}, and hence, its performance heavily depends on the maximal neutrino energy in the system, $E_{\nu,\text{max}}$. The scaling of the computational time is 
\begin{equation}
t_{\text{computation}}\propto E_{\nu,\text{max}}^{4},
\end{equation}
Here, one power of $E_{\nu,\text{max}}$ comes from the scaling of the number of time steps required to resolve the thermalization of neutrinos, another power is due to the number of bins (scaling as $\propto E_{\nu,\text{max}}$), and two powers are due to the dimensionality of the analytically reduced collision integral (see ref.~\cite{Ovchynnikov:2024rfu} for more discussions). In the case of new physics decaying into muons and pions, when $E_{\nu,\text{max}}\approx m_{\mu}/2$, the computational time needed to obtain $N_{\text{eff}}$ for the single LLP setup is $\mathcal{O}(1\text{ hour})$. However, it increases by orders of magnitude if, e.g., kaons are present (as then $E_{\nu,\text{max}}\approx m_{K}/2$), making the solver inapplicable in practice.

Therefore, in this study, we restrict ourselves to scenarios of LLPs decaying into pions, muons, or neutrinos with energies $E_{\nu}\lesssim 100\text{ MeV}$. The cases of decays into kaons, as well as neutrinos with higher energies, may be studied using different approaches, such as the Direct Simulation Monte Carlo from refs.~\cite{Ovchynnikov:2024rfu,Ovchynnikov:2024xyd}.

\section{Derivation of the source term in the unintegrated Boltzmann equations}
\label{app:source_BE}

In this Appendix, we derive the source term for neutrinos, eq.~\eqref{eq:source-discretization}, i.e., the collision term from metastable particle decays to neutrinos, in the unintegrated Boltzmann equations. 

We consider the decay process 
\begin{equation}
X\rightarrow Y\rightarrow \nu_\alpha+\dots,
\end{equation}
where, as usual, $X$ denotes the LLP, while $Y$ stays for metastable SM particles decaying into neutrinos. We can easily generalize the method to include other decay processes such as $X\rightarrow Y\rightarrow Z\rightarrow \nu_\alpha+...$ and $X\rightarrow \nu_\alpha+...$, etc.

The source term for $Y \rightarrow \nu_\alpha+...$ in the neutrino Boltzmann equations is 
\begin{align}
    \left(\frac{df_{\nu_\alpha}}{dt}\right)_{\rm source}=\langle P_{\alpha\beta}\rangle I_{Y\rightarrow \nu_\beta+...},
\end{align}
where $I_{Y\rightarrow \nu_\beta+...}$ is the corresponding collision term,
\begin{align}
    I_{Y\rightarrow \nu_\beta+...}=\frac{1}{2E_{\nu_\beta}}\int \frac{d^3p_Y}{(2\pi)^32E_Y}\prod_{f=2}\frac{d^3p_f}{(2\pi)^32E_f}S|\mathcal{M}_{Y\rightarrow_{\nu_\beta}+...}|^2(2\pi)^3\delta^{(4)}(p_Y-p_{\nu_\beta}-...)f_Y(p_Y),
\end{align}
where $f=1$ corresponds $\nu_\beta$, $f_Y(p_Y)$ is the distribution function for $Y$, and we neglect the Pauli-blocking/Bose-enhancement factors. $Y$ becomes non-relativistic before their decay due to the strong EM scattering. In this case, the collision term is
\begin{align}
    I_{Y\rightarrow \nu_\beta+...}=\frac{n_Y}{2E_{\nu_\beta}2m_Y}\int\prod_{f=2}\frac{d^3p_f}{(2\pi)^32E_f}S|\mathcal{M}_{Y\rightarrow_{\nu_\beta}+...}|^2(2\pi)^3\delta^{(4)}(m_Y-p_{\nu_\beta}-...).
\end{align}
The decay rate for $Y\rightarrow \nu_\beta+...$ at the rest frame of $Y$ is 
\begin{align}
    \Gamma_{Y}=\frac{1}{2m_Y}\int\frac{d^3p_{\nu_\beta}}{(2\pi)^32E_{\nu_\beta}}\prod_{f=2}\frac{d^3p_f}{(2\pi)^32E_f}S|\mathcal{M}_{Y\rightarrow_{\nu_\beta}+...}|^2(2\pi)^3\delta^{(4)}(m_Y-p_{\nu_\beta}-...).
\end{align}
The collision term is rewritten in terms of $d\Gamma_{Y}/dp_{\nu_\beta},\ \Gamma_{Y}$ and $\tau_Y= \Gamma_{Y}^{-1}$,
\begin{align}
    I_{Y\rightarrow \nu_\beta+...}=\frac{n_Y}{\tau_Y}\frac{1}{\Gamma_Y}\frac{d\Gamma_Y}{dp_{\nu_\beta}}\frac{2\pi^2}{p_{\nu_\beta}^2},
\end{align}
where $n_Y/\tau_Y$ is the production rate of neutrinos from Y decays and
$\frac{1}{\Gamma_Y}\frac{d\Gamma_Y}{dp_{\nu_\beta}}\equiv \mathcal{F}_{\nu_\beta}(p_{\nu_\beta})$ is the momentum distribution (normalized to be unity) for a neutrino by a decay. The number density for $Y$ produced by the $X$ decay is given in eq.~\eqref{eq:Yabundance},
\begin{align}
    n_Y(t)=n_X(t)N_Y^X\frac{\tau_Y}{\tau_X}P_{\rm decay}^Y(t),
\end{align}
where $n_X(t)$ is given by eq.~\eqref{eq:nX}.
Finally we obtain the source term for $X\rightarrow Y\rightarrow \nu_\beta+...$,
\begin{align}
    \left(\frac{df_{\nu_\alpha}(p)}{dt}\right)_{\rm source}=\langle P_{\alpha\beta}\rangle \times\frac{n_X(t)}{\tau_X}N_Y^X\mathcal{F}_{\nu_\beta}(p)P_{\rm decay}^Y(t),
\end{align}
where $\frac{n_X(t)}{\tau_X}N_Y^X$ is the production rate of $Y$ from $X$ decays.
Following sec.~\ref{sec:metastable-dynamics}, we obtain the decay probability for $Y$, $P_{\rm decay}^Y$, and then we can compute the source term. Generalizing the decay processes, we obtain the source term for the pion decay to neutrinos~\eqref{eq:source-discretization}.

\bibliography{bib.bib}

\end{document}